\input harvmac
\input epsf
%
\def\abstract#1{
\vskip .5in\vfil\centerline
{\bf Abstract}\penalty1000
{{\smallskip\ifx\answ\bigans\leftskip 2pc \rightskip 2pc
\else\leftskip 5pc \rightskip 5pc\fi
\noindent\abstractfont \baselineskip=12pt
{#1} \smallskip}}
\penalty-1000}
\def\us#1{\underline{#1}}
\def\hth/#1#2#3#4#5#6#7{{\tt hep-th/#1#2#3#4#5#6#7}}
\def\nup#1({Nucl.\ Phys.\ $\us {B#1}$\ (}
\def\plt#1({Phys.\ Lett.\ $\us  {B#1}$\ (}
\def\cmp#1({Comm.\ Math.\ Phys.\ $\us  {#1}$\ (}
\def\prp#1({Phys.\ Rep.\ $\us  {#1}$\ (}
\def\prl#1({Phys.\ Rev.\ Lett.\ $\us  {#1}$\ (}
\def\prv#1({Phys.\ Rev.\ $\us  {#1}$\ (}
\def\mpl#1({Mod.\ Phys.\ Let.\ $\us  {A#1}$\ (}
\def\ijmp#1({Int.\ J.\ Mod.\ Phys.\ $\us{A#1}$\ (}
\def\ul#1{$\underline{#1}$}

\def\IZ{{\bf Z}}

\def\IR{\relax{\rm I\kern-.18em R}}

\def\al{\alpha}

\def\al{\alpha}

\def\tr{{\rm tr}}
\def\Tr{{\rm Tr}}

\def\ul#1{\underline{#1}}

\def\DBox#1{\mathop{\mkern0.5\thinmuskip
           \vbox{\hrule\hbox{\vrule\hskip#1\vrule height#1 width 0pt\vrule}
           \hrule
           \hbox{\vrule\hskip#1\vrule height#1 width 0pt\vrule}
           \hrule}\mkern0.5\thinmuskip}}
\def\Dbox{{\DBox{2pt}}}

\def\TBox#1{\mathop{\mkern0.5\thinmuskip
           \vbox{\hrule\hbox{\vrule\hskip#1\vrule height#1 width 0pt\vrule}
           \hrule
           \hbox{\vrule\hskip#1\vrule height#1 width 0pt\vrule}
           \hrule
           \hbox{\vrule\hskip#1\vrule height#1 width 0pt\vrule}
           \hrule}\mkern0.5\thinmuskip}}
\def\Tbox{{\TBox{2pt}}}

\def\sss{\scriptscriptstyle}

\vbadness=10000

\parindent10pt

\lref\AFIQ{G. Aldazabal, A. Font, L. Ibanez
and F. Quevedo, ``Chains of N=2, D=4 heterotic/type II duals'',
Nucl.\ Phys.\ B461 (1996) 85, hep-th/9510093}
\lref\HW{P.~Ho\v{r}ava and E.~Witten,
``Eleven-dimensional supergravity on a manifold with boundary'',
Nucl.\ Phys.\  B475 (1996) 94,
hep-th/9603142; ``Heterotic and type I string dynamics 
from eleven-dimensions'',
Nucl.\ Phys.\ B460 (1996) 506,
hep-th/9510209}
\lref\SW{N. Seiberg and E. Witten, ``Comments on String Dynamics
in Six Dimensions'', Nucl.\ Phys.\ B471 (1996) 121, hep-th/9603003}
\lref\stieberger{S. Stieberger, ``(0,2) Heterotic Gauge Couplings
and their M-Theory Origin'', Nucl.\ Phys.\ B541 (1999) 109, hep-th/9807124}
\lref\DMW{M. Duff, R. Minasian and E. Witten,
``Evidence for Heterotic/Heterotic Duality'', Nucl. Phys. B465 (1996) 413,
hep-th/9601036}
\lref\FOL{M. Faux, D. L\"ust and B. Ovrut, ``Intersecting Orbifold
Planes and Local Anomaly Cancellation in M-Theory'',
Nucl.\ Phys.\ B554 (1999) 437, hep-th/9903028}
\lref\KLT{V.~Kaplunovsky, J.~Louis and S.~Theisen,
``Aspects of duality in N=2 string vacua'',
Phys.\ Lett.\ {\bf B357} (1995) 71,
hep-th/9506110.}
\lref\AGW{ L. Alvarez-Gaume and E. Witten, ``Gravitational anomalies'',
Nucl. Phys. B234 (1984) 269}
\lref\GSW{M. Green, J. Schwarz and E. Witten, Superstring Theory,
Vol II, CUP 1987}
\lref\BLPSSW{M.~Berkooz, R.G.~Leigh, J.~Polchinski, 
J.H.~Schwarz, N.~Seiberg and E.~Witten,
``Anomalies, Dualities, and Topology of D=6 N=1 Superstring Vacua'',
Nucl.\ Phys.\ {B475}, 115 (1996)
hep-th/9605184}
\lref\DLM{M. Duff, J. Liu and R. Minasian, ``Eleven Dimensional Origin
of String/String Duality: a One-Loop Test'', 
Nucl.\ Phys.\ B452 (1995) 261, hep-th/9506126}
\lref\Witten{E. Witten, ``String Dynamics in Various Dimensions'', 
Nucl.\ Phys.\ B443(1995) 85 hep-th/9503124}
\lref\Sen{
A. Sen, ``Orbifolds of M-Theory and String theory'', 
Mod.\ Phys. Lett. A11 (1996) 1339, hep-th/9603113}
\lref\AFIUV{G. Aldazabar, A. Font, L. Ibanez, A. Uranga and G. Violero,
``Non-Perturbative Heterotic D=6, N=1 Orbifold Vacua'', 
Nucl.\ Phys.\ B519 (1998) 239, hep-th/9706158}
\lref\Senrev{A.~Sen,
``An introduction to non-perturbative string theory'',
hep-th/9802051}
\lref\Townsendrev{P.K.~Townsend,
``Four lectures on M-theory'',
hep-th/9612121}
\lref\Erler{J.~Erler,
``Anomaly cancellation in six-dimensions'',
J.\ Math.\ Phys.\  {35}, 1819 (1994),
hep-th/9304104}

\Title{\vbox{
\rightline{\vbox{\baselineskip12pt
\hbox{UTTG--08--99}
\hbox{TAUP-2606-99}
\hbox{hep-th/9912144}}}}}
{On the Duality between Perturbative Heterotic}
\vskip-1cm\centerline{{\titlefont Orbifolds and M-Theory on $T^4/Z_N$
\footnote{$^{\scriptscriptstyle*}$}{\sevenrm Research supported in part by
the US-Israeli Binational Science Foundation, the US National Science
Foundation (grant $\#$PHY--95--11632), the Robert A.~Welsh Foundation,
the German--Israeli Foundation for Scientific Research (GIF),
by the European Commission TMR programme ERBFMRX--CT96--0045,
the Israel Science Foundation and by DFG--SFB--375.
}}}
\vskip 0.3cm
\centerline{V. Kaplunovsky$^a$, J. Sonnenschein$^b$, S. Theisen$^c$,
S. Yankielowicz$^b$}
\vskip 0.6cm
\centerline{$^a$ \it Theory Group, Dept. of Physics,
University of Texas, Austin, TX 78712, USA}
\vskip 0.0cm
\centerline{$^b$ \it School of Physics and Astronomy, Beverly and
Raimond Sackler Faculty of Exact Sciences,}
\vskip-,1cm
\centerline{\it Tel Aviv University,
Ramat-Aviv, Tel-Aviv 69978, Israel}
\vskip 0.0cm
\centerline{$^c$ \it Sektion Physik, Universit\"at M\"unchen,
Theresienstrasse 37, D-80333 M\"unchen, Germany}

\abstract{
The heterotic $E_8\times E_8$ string compactified on an orbifold $T^4/\IZ_N$
has gauge group $G\times G'$ with (massless) states in its twisted
sectors which are charged under both gauge group factors. In the
dual M-theory on $(T^4/\IZ_N)\otimes(S^1/\IZ_2)$ the two group factors are
separated in the eleventh direction and the $G$ and $G'$ gauge fields
are confined to the two boundary planes, respectively. We present a
scenario which allows for a resolution of this apparent paradox
and assigns all massless matter multiplets locally to
the different six-dimensional
boundary fixed planes. The resolution consists of diagonal
mixing between the gauge groups which live on
the connecting seven-planes (6d and the eleventh dimension)
and one of the gauge group factors. We present evidence supporting
this mixing by considering gauge couplings and verify local
anomaly cancellation. We also discuss open problems which arise in the
presence of $U_1$ factors.
}
\Date{\vbox{\hbox{\sl {December 1999}}
}}
\goodbreak

\parskip=4pt plus 15pt minus 1pt
\baselineskip=15pt plus 2pt minus 1pt

\newsec{Introduction}
The r\^ole of M-theory for string duality is undisputed.
Nevertheless, we are far from understanding this 11-dimensional
theory at a fundamental level. We know that at particular points in its
moduli space all known string theories are recovered. Moreover,
the low energy physics is captured by 11-dimensional supergravity.
Compactification and duality symmetries provide data which allow
us to gain some insight into the structure of M-theory 
(see {\it e.g.} refs. \Witten, \Senrev, \Townsendrev\ for reviews).

In this paper we concentrate on 10-dimensional $E_8\times E_8$
heterotic string theory compactified on $T^4/\IZ_N$ orbifolds
and its dual M-theory description. The $E_8\times E_8$ heterotic
string is dual to M-theory on $S^1/\IZ_2\simeq I$. Ho\v{r}ava and
Witten have shown \HW\ that one $E_8$ factor lives on each of the two
boundary ten-planes\foot{Throughout we refer to an extended object as
a {\sl plane}. {\it E.g.} a ten-plane has ten space-time dimensions.}
and that the corresponding gauge multiplets are confined to it.
There is no a priori M-theoretical explanation of the appearance
of the $E_8$ gauge group on each of the two ten-planes,
but they are needed for local anomaly cancellation.
Thus, anomaly cancellation provides important constraints which teach us
about the structure of M-theory.

Considering the proposed duality between the heterotic string on
$\IR^{5,1}\otimes(T^4/\IZ_N)$ and M-theory on
$\IR^{5,1}\otimes (T^4/\IZ_N)\otimes(S^1/\IZ_2)$ one immediately
encounters the following puzzle.
Suppose both ten-plane $E_8$ groups are broken in the six-dimensional theory;
let $G\times G'\subset E_8\times E_8$ denote the surviving subgroup.
The twisted sectors of heterotic orbifolds generally contain
massless states which are charged under {\sl both $G$ and $G'$}.
In the Ho\v{r}ava--Witten theory however, $G$ is $G$ and $G'$ is $G'$, they
live on different ten-plains and nowhere the twain shall meet,
so there does not seem to be any place where a massless state can be
simultaneously charged with respect to both $G$ and $G'$.
Indeed, how would a state residing at one end of the eleventh dimension
know about the gauge group acting on the other side?
Somehow, in the effective seven-dimensional gauge theory on 
$\IR^{5,1}\otimes(S^1/\IZ_2)$ gauge quantum numbers ought to `flow'
from one end of the $x^{11}$ to the other end;
the main objective of our work is to understand how this works. 

The same problem also arises in the phenomenologically more interesting
examples of orbifold compactifications to four dimensions.
In fact, our initial motivation was to understand how the gauge quantum
numbers work in the four-dimensional orbifolds, but the situation
in six dimensions turned out to be easier.
In particular, we got very useful hints from the requirement of
local anomaly cancellation, which are much stronger in $d=6$ than in $d=4$.
Consequently, in this article we restrict ourselves to the discussion of the
six-dimensional theories with the intention 
to come back to the four-dimensional case in the future.

In M-theory, an $A_{n-1}$ singularity of the K3 compactification
--- such as a $\IZ_n$ fixed point of an orbifold
--- supports an $SU_n$ gauge theory on the corresponding seven-plane
$\IR^{5,1}\otimes\{{\rm f.p.}\}\otimes(S^1/\IZ_2)$.
The Cartan sub-algebra
arises from the zero modes of the three-form potential of 11d SUGRA
and the completion to $SU_n$ is achieved by M2 branes wrapped on a
vanishing 2-cycle in the orbifold limit of the smooth K3 compactification
manifold.
These seven-plane gauge groups play a vital r\^ole in our resolution
of the paradoxical $G\times G'$ charges of the twisted states:
At one end of the eleventh dimension, say at $x^{11}=0$, the seven-plane
$SU_n$ {\it mixes} with a similar factor of $G$ --- the unbroken subgroup
of the $E_8$ living on the ten-plane boundary of the entire 11d spacetime.
The mixing happens along the six-planes where the fixed seven-planes intersect
the $x^{11}=0$ ten-plane, but it has global consequences for the resulting
effective theory:
The {\sl diagonal} ten-plane/seven-plane $SU_n$ gauge group {\sl appears}
to be a subgroup of the ten-plane $G$, but actually reaches
along the fixed seven-planes to the other end of the~$x^{11}$.
Consequently, along the six-planes where the fixed seven-planes carrying
the $SU_n$ intersect the second ten-plane carrying $G'$, we have both
$SU_n$ and $G'$ gauge fields at the same location in space ---
and hence the twisted states living on those six-planes may
have both the $SU_n^{\rm diag}$ and the $G'$ charges in a perfectly local fashion.

The bottom line is, the twisted states have local $G'\times SU_n^{7P}$
charges but from the global point of view, they have
simultaneous charges under the $G'$ and the diagonal $SU_n$ gauge groups.
The apparent paradox of simultaneous $G'$ and $G$ charges is due
to mis-identification of the diagonal ten-plane/seven-plane $SU_n$
as a subgroup of the ten-plane $G\subset E_8$.
This mis-identification is natural in the perturbative heterotic string theory
where the entire eleventh dimension is invisible and everything lives in
ten dimensions.
In the M-theory, one needs to be more careful.

In this article, we shall marshal three lines of evidence
 for the mixing of ten-plane and seven-plane gauge groups.
First, this is the only way to reconcile the massless spectra of heterotic
orbifolds with locality of the dual M-theory description.
Second, the heterotic gauge couplings
(which can be computed exactly in six dimensions)
will show that some $SU_n$ gauge groups cannot be of purely perturbative origin
but must be diagonally mixed with several non-perturbative factors,
\eqn\GaugeMixing{
SU_n^{d=6}\ =\ \mathop{\rm diag}\left[ SU_n^{\rm pert}\times
\left(SU_n^{\rm non-pert}\right)^\nu \right] ,
}
and the number $\nu$ of the non-perturbative $SU_n$ factors will always turn out
to be equal to the number of fixed seven-planes in M-theory that carry $SU_n$
gauge groups.
Finally, each six-plane in M-theory suffers from local anomalies
which are sensitive to spectra of massless particles living on the six-planes
themselves, on the seven-planes, on the ten-planes and in the eleven-dimensional
bulk as well as inflow and intersection anomalies due to the M-theory's
Chern--Simons terms.
In six dimensions, it is very difficult to cancel the local anomalies unless
one has correct {\sl local} spectra of all the fields --- and we shall see that
the ten-plane/seven-plane gauge group mixing indeed provides for cancellation
of all the local anomalies.

The rest of this paper is organized as follows:
We begin with an illustrative example of a $\IZ_2$ orbifold.
In section~2 we take a close look at the seven-plane and the ten-plane
$SU_2$ gauge groups of this orbifold and discuss their mixing from
both heterotic and M-theory points of view.
In particular, we show how the mixing explains the $SU_2$ charges of
the twisted states as well as the exact value of the $SU_2$ gauge coupling.
In section~3 we confirm our solution by verifying local anomaly cancellation.
In these two sections we try to be as explicit as possible.
Section~4 generalizes our approach to other $\IZ_N$ orbifolds.
The subtleties that arise in non-prime orbifolds are treated in detail.

Unfortunately, our proposed solution works for some $\IZ_N$ orbifolds
but has difficulties with others.
Section~5 describes two common types of complications, both associated
with broken seven-plane $SU_n$ groups.
In some orbifolds (discussed in section~5.1), the perturbative ten-plane
gauge groups don't mix with the non-perturbative seven-plane groups.
As far as the effective six-dimensional effective theory is concerned, the
seven-plane groups are completely invisible, but the local anomalies on the
six-planes are not so blind.
To cancel the anomalies, we have to assume that the seven-plane gauge groups
are not $SU_n$'s but rather their Cartan subgroups $U_1^{(n-1)}$.
Alas, from the seven-dimensional point of view, all $\IZ_n$ fixed planes
are created equal and we have no idea how or why does the M-theory decide
that such fixed planes carry full $SU_n$ gauge groups in some orbifolds but
only the Cartan $U_1^{(n-1)}$ subgroups in others.
Worse problems plague orbifold models where the ten-plane and the seven-plane
gauge groups do mix but the mixing involves abelian factors.
In section 5.2 we show that in such models local anomaly
cancellation does not seem to work
and we speculate how the seven-dimensional Chern--Simons terms might
remedy this problem.

Section~6 summarizes our results.
Appendices A, B and C contain some useful data about the anomalies in
six dimensions and as well as some related group theory.

\newsec{Compactification on $\IZ_2$}

\subsec{Heterotic vs. M-theory point of view}

Consider the perturbative
compactification of the $E_8\times E_8$ heterotic string
on the $T^4/\IZ_2$ orbifold limit of K3.
$\IZ_2$ acts as
$(z^1,z^2)\to -(z^1,z^2)$ on the two complex
coordinates of $T^4$. Under this transformation all
four moduli of the torus are invariant and are thus also
moduli of the orbifold. There are sixteen orbifold fixed points.
This compactification has ${\cal N}=1$ supersymmetry
in $d=6$ (eight supercharges).
We represent the discrete $\IZ_2$ transformation in the
$E_8\times E_8$ gauge lattice via the shift vector
$\delta=({1\over2},{1\over2},0,0,0,0,0,0;1,0,0,0,0,0,0,0)$. This results
in the gauge group
$G\times G'=[E_7\times SU_2]\times SO_{16}\subset E_8\times E_8$.
The massless matter in the untwisted sector consists of hyper-multiplets
transforming as $(\ul{56},\ul{2};\ul{1})$ and
$(\ul{1},\ul{1};\ul{128})$. They only carry charges of $G$ or $G'$.
The untwisted sector also includes the four moduli which are
gauge singlets.
The massless  matter in the twisted sector
consists of sixteen half-hyper-multiplets,
one localized at each fixed point, transforming as $(\ul{1},\ul{2};\ul{16})$.
They carry quantum numbers under {\it both},
$G$ {\it and} $G'$. This is the complete massless matter spectrum of this
compactification. The rules to determine the massless states of heterotic
K3 orbifold compactifications have recently been reviewed in \AFIQ.
The spectra of some of the models considered in this paper were 
constructed in \Erler.
We note that the difference of the number
of hyper-multiplets ($n_H$) and of vector-multiplets ($n_V$) satisfies
$n_H-n_V=244$, as required for a consistent perturbative heterotic
compactification.

We want to study this compactification within the context of the
conjectured duality between the heterotic string on
K3 and M-theory on ${\rm K3}\otimes(S^1/\IZ_2)\simeq {\rm K3}\otimes I$.
In the latter description,
the gauge fields of $G\times G'$ are confined to one of the two
ten-planes at the ends of the $x^{11}$ interval. We will denote them by
10P and 10P$'$, respectively.
Since none of the perturbative gauge fields live in the bulk,
it is therefore not a priori clear
how the twisted matter fields, which are charged under $G$ and
$G'$, can be accommodated in the M-theory picture.
{}From the six-dimensional point of view each of the sixteen
fixed points of the heterotic compactification
$T^4/\IZ_2$ is a fixed six-plane.
In the M-theory picture there are sixteen
seven-planes, denoted 
by 7P, of infinite extent in six space-time directions
and of finite extent in the $x^{11}$ direction, $x^{11}\in I=[0,\pi R_{11}]$.
Their boundaries are six-planes which are the intersection of the 7P with the 
two ten-planes at $x^{11}=0$ and $x^{11}=\pi R_{11}$.
It is here where the perturbative
heterotic gauge groups are located, $G$ say at $x^{11}=0$ and
$G'$ at $x^{11}=\pi R_{11}$. The sixteen intersection six-planes on the
$E_7\times SU_2$ side will be denoted as I6 and those on the
$SO_{16}$ side by I6$'$. All the I6 are of course completely equivalent,
as are the I6$'$.

{}From this geometric picture it is reasonable to expect that the 7P's
do play a central role in the resolution of the puzzle with
the massless twisted heterotic states in the
M-theory context. They are the only objects which connect both
sides of the $x^{11}$ interval.\foot{Our discussion is restricted to
massless states. Their masslessness is protected by their chirality.
There are massive states which are charged under both $E_7$ and $SO_{16}$.
Since the ${\cal N}=1$ SUSY algebra in $d=6$ has no central charge,
they are not BPS and their masses are not protected against perturbative and
non-perturbative corrections. One may speculate that these states originate
from open M2 branes stretched between the two ten-planes \HW.
{}From now on we only consider massless states. }
On their world-volume we have a
seven-dimensional supersymmetric gauge theory.

{}For each $\IZ_2$ fixed point of the orbifold
there is an associated harmonic two-form,
the K\"ahler form of the $S^2$ which has shrunk to zero size at the
orbifold singularity.
In M-theory on $T^4/\IZ_2$ the
three-form potential $C$ of eleven-dimensional supergravity
with two internal indices and one space-time index
has thus a zero mode associated with each orbifold fixed-point.
In other words,
there is a $U_1$ vector, and by supersymmetry, a complete
seven-dimensional vector-multiplet for each fixed point.\foot{Since
$H^2({\rm K3})=22$, there are six additional vector-multiplets which are
not attached to an orbifold singularity. Upon
compactification, their components arrange themselves, together
with components of the metric, to the supergravity multiplet,
a tensor multiplet and four moduli hyper-multiplets.}
Wrapping M2 branes around the $S^2$ gives rise to additional
massless states in the limit of zero volume of the $S^2$.
Taking into account two possible orientations gives $SU_2$ as the
maximal gauge group.
We now compactify further to six dimensions on $S^1/\IZ_2$. This breaks
half the supersymmetry and each seven-dimensional vector-multiplet
decomposes into a six-dimensional vector-multiplet $V_7$
and a hyper-multiplet $H_7$ (the subscript reminds of their
seven-dimensional origin).
If the five-dimensional compact manifold is the direct
product $(T^4/\IZ_2)\otimes (S^1/\IZ_2)$, the vector components
which arise from $C$ are projected out. This is because
$C$ is odd under $\IZ_2:\,x^{11}\to - x^{11}$ \HW. 
In particular, the
vector components must vanish on the intersection six-planes.
There is no such restriction on the hyper-multiplet components.

The analysis of this model which we 
present in sects.~2 and 3 as well as the analysis of other models 
in sect.~4 requires some modification
of the set-up such that we retain vectors of the
non-perturbative gauge group.
In fact we will argue that the vector-components of the full
non-perturbative $SU_2$ survives on I6$'$ and the hyper component
on I6. This clearly requires a departure from the direct product
geometry assumed above to a `twisted product'. We must admit that
we do not know how this works in detail. We believe that
the proposed field content on any of the planes involved, in particular on the
7P and the I6 and I6$'$, is correct. We give several
pieces of strong evidence. Rather than being able to specify how exactly the
twist acts, we can only state the effect it has on the boundary conditions
of the $H_7$ and $V_7$ fields. To understand what is meant here, recall
that the presence of the ten-planes,
or in other words, dividing by $\IZ_2$, breaks half of the
supersymmetries, 8 of the 16 supercharges which would be present in
M-theory on K3 are even and 8 are odd under the $\IZ_2$.
This entails that under the $7d\to 6d$ decomposition
the vector and the hyper components of the
seven-dimensional vector-multiplet have opposite
(free vs. fixed) boundary conditions at each end of the interval.
The choice of the boundary conditions on both sides will be crucial
below.
As we shall see, we have to impose Dirichlet boundary conditions
for the $SU_2$ vector components on the I6$'$ and Neumann conditions
on the I6. This is to be compared with the $U_1$ vectors which
are not associated to orbifold fixed points (see the previous footnote).
Here the boundary conditions are such that the zero mode for their
hyper components are retained. The vector components are projected out by
the $\IZ_2$ twist.
We want to stress once more that we are not able to derive these
boundary conditions from first principles but we will present compelling
arguments in favour of them.

Now that we have introduced the main ingredients of the
model from the M-theory point of view, we
can give a qualitative description of how they are assembled
into a picture that is consistent with the heterotic
description. This will involve the non-perturbative
gauge groups in an essential way and we will in fact establish that
the $SU_2$ visible in the heterotic description is the
diagonal subgroup
$SU_2^{\rm het}={\rm diag}[SU_2^{\rm pert}
\times\left(SU_2^{\rm non-pert}\right){}^{\!\!16}]$.

By looking at the heterotic spectrum of the model,
it is clear that the charged states in the untwisted sector
live on the ten-planes,
($\ul{56},\ul{2};\ul{1}$) on 10P and
($\ul{1},\ul{1};\ul{128}$) on 10P$'$. The major new ingredient
in the M-theory description of the model is the presence of additional
gauge groups, the non-perturbative
$SU_2$'s, one on each 7P, {\it i.e.} one for each orbifold fixed-point.
This means that we have to reconsider the $SU_2$ charge assignments
of the fields in the twisted sector. The
$E_7$ and the $SO_{16}$ gauge factors are unaffected by the
presence of the non-perturbative $SU_2$'s.
The twisted matter fields necessarily live either on
the I6 or I6$'$ intersection planes. As they are charged under the $SO_{16}$,
which is confined to 10P$'$, 
they are located on the I6$'$'s, one half-hyper-multiplet on each.
It then also follows that we have to attribute their $SU_2$
quantum numbers to the non-perturbative $SU_2$ which lives on the 7P
which is bounded by the I6$'$.
The situation is illustrated in the figure.
\vskip 1cm
\epsfbox{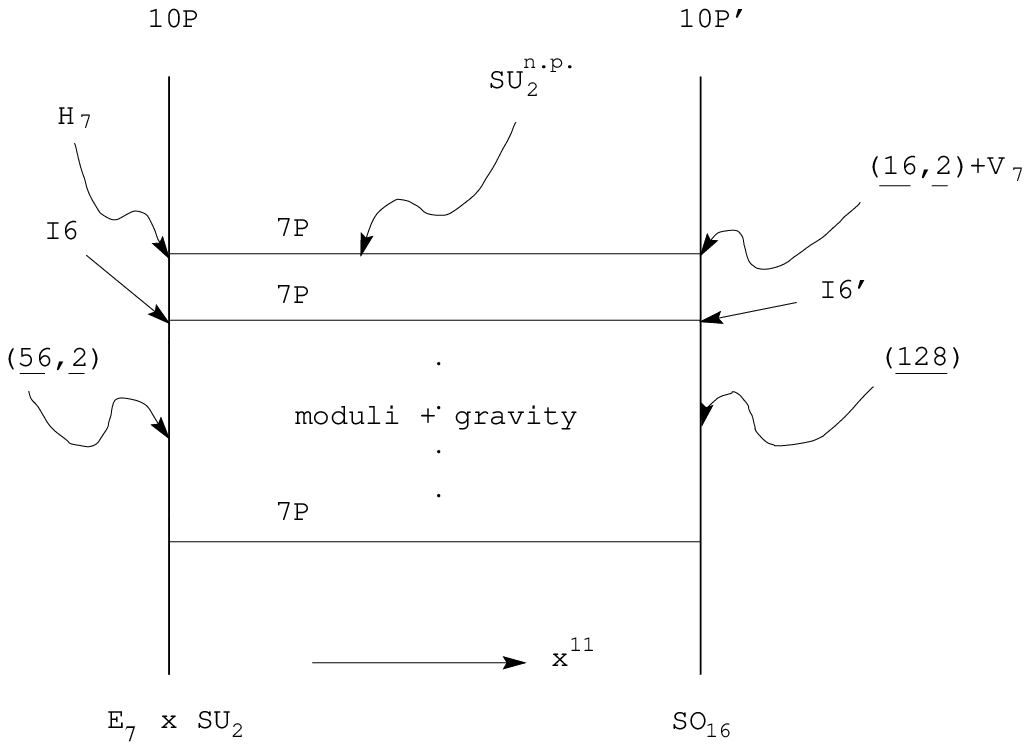}
\vskip 1cm
We conclude that the 
$i{\rm th\over{}}$ twisted matter multiplet transforms as
$(\ul{16},\ul{2})$ of 
$SO_{16}^{(\rm pert)}|_{10P'}\times SU_2^{(\rm non-pert)}|_{7P_i}$,
where $i=1,\dots,16$. In particular it does not couple to the
perturbative $SU_2$. For this picture to be consistent, we have
to impose adequate boundary conditions for $H_7$ and $V_7$.
The twisted matter multiplet can couple to the gauge field only
if we impose free (Neumann) boundary condition on $V_7$ at the
I6$'$ end of the 7P. As explained before, this implies fixed
boundary conditions for $H_7$, which is thus invisible at I6$'$.
We now have to cope with the fact that
in the heterotic picture there is only one $SU_2$ gauge factor.
This will be consistent with the M-theory description if
the perturbative $SU_2$ which is confined to 10P
mixes with the sixteen non-perturbative $SU_2$'s such that the
heterotic $SU_2$ is the diagonal subgroup
$SU_2^{\rm het}={\rm diag}[SU_2^{\rm pert}
\times \left(SU_2^{\rm non-pert}\right){}^{\!\!16}]$.
This requires that the $SU_2$
vector-multiplets $V_7$ are
locked to the perturbative $SU_2$ on 10P. That is to say that we have
to impose
\eqn\lock{
A_\mu^{\rm non-pert}(x^1,\dots ,x^6,x^{11}\!=\!0)=
A_\mu^{\rm pert}(x^1,\dots,x^6,x^7\!=\!x^8\!=\!x^9\!=\!x^{10}\!=\!0)\,,
\quad\hbox{for}\,\,\mu=1,\dots,6,}
and likewise for the gauginos and at all other fixed points.
Imposing fixed boundary conditions for $V_7$ on 10P requires
free boundary conditions for
the adjoint hyper-multiplet $H_7$. Thus
the latter is visible on the I$6$. This is also indicated
in the figure.

In the following we will substantiate this picture in two independent
ways. In sect 2.2 we consider the heterotic gauge couplings and
we will find that indeed the $SU_2$ coupling has both a perturbative
contribution, which is due to $SU_2|_{10P}$ and a non-perturbative
contribution which we interpret to arise from the sixteen
$SU_2|_{7P}$'s. In sect.~2.3 we then show that the local anomalies
on all intersection six-planes cancel. In particular this
includes quantum contributions from either $V_7$ or $H_7$ and
also inflow contributions from the bulk and the seven-planes.
This is another check on the
assignment of the fields to the different planes, as summarized in the
figure.

\subsec{Consideration of the gauge couplings}

The gauge kinetic energy of the six-dimensional
low-energy effective ${\cal N}=1$
SYM theory is, in string frame, up to a numerical constant \DMW
\eqn\gaugekin{
{\cal L}\sim{1\over\al'}\sum_{\al}(v_\al e^{-\phi}+\tilde v_\al)\tr F_\al^2\,.}
Here $\phi$ is the unique dilaton of the
perturbative heterotic string. Additional dilatons arise if we allow
for additional tensor-multiplets, but this we will not do
in this paper. $e^{-\phi}\sim {{\rm Vol(K3)}\over \lambda_H^2\al'^2}$,
where $\lambda_H$ is the heterotic coupling constant.
The sum is over all gauge group factors.
$v$ and $\tilde v$ are dimensionless constants.
{}For perturbative gauge groups, $v=1$ -- it is in fact the
level of the Kac-Moody algebra -- and $\tilde v$ arises at one loop.
For non-perturbative gauge groups, on the other hand, $v=0$
and $\tilde v$ is fixed at tree level.
The coefficients $v$ and $\tilde v$ are related, via supersymmetry,
to the coefficients of the anomaly polynomial which must factorize
to allow a Green-Schwarz mechanism to cancel the anomaly.
For further explanation we refer to Appendix~B.
Factorizability of the anomaly polynomial imposes the constraint
\eqn\bv{
b_\al=6(v_\al+\tilde v_\al)}
where $b_\al$ is the coefficient of the one-loop beta-function of the
$d=4$ SYM theory that one obtains upon further compactification
on $T^2$.
Given the matter content of the theory we can thus compute $\tilde v_\al$.

The Ho\v{r}ava--Witten theory, at least when applied to compactifications on a
{\it smooth} K3, relates $\tilde v$ to the net magnetic charge $k$
on the ten-plane on which it lives. The field strength $G$ of the
three-from potential $C$ satisfies \DMW\
\eqn\smooth{
dG={1\over16\pi^2}\Bigl\lbrace\delta(x^{11})
\Bigl({1\over2}\tr R^2-\tr F_1\wedge F_1\Bigr)
+\delta(x^{11}\!\!-\!\pi R_{11})\Bigl({1\over2}\tr R^2-\tr F_2\wedge F_2\Bigr)
\Bigr\rbrace dx^{11}}
from which the magnetic charge of the
ten-planes is determined as $k_{1,2}=n_{1,2}\!-\!12$.
$n_{1,2}={1\over 16\pi^2}\int_{K3}\tr F_{1,2}^2$
are the instanton numbers on the two sides of the interval. 
In perturbative compactifications, {\it i.e.} in the absence of
freely floating M5 branes, $n_1+n_2=24$ and therefore $k_1+k_2=0$.
Also, integrating the anomaly polynomial of the ten-dimensional
heterotic theory over a smooth K3 one derives
$\tilde v_{1,2}={1\over 2}k_{1,2}$ and thus $\tilde v_1+\tilde v_2=0$.
In particular, if two group factors arise from the same $E_8$, they
should have the same $\tilde v$.
{}For the model at hand, $n_1=8$ and $n_2=16$. Therefore, if the above
analysis applied, we would find $\tilde v=+2$ for the
$SO_{16}$ factor and $\tilde v=-2$ for both the $E_7$ and the $SU_2$ factor.
However, using \bv\ and the field content of the theory, we find
\eqn\bes{\eqalign{
b(E_7)=-6\quad & \rightarrow \quad\tilde v(E_7)=-2\,,\cr
b(SU_2)=90\quad & \rightarrow \quad\tilde v(SU_2)=14\,,\cr
b(SO_{16})=18\quad & \rightarrow \quad\tilde v(SO_{16})=2\,.}}
We thus realize that for the orbifold compactification the Ho\v{r}ava--Witten
formulae work for the $E_7$ and the $SO_{16}$ coupling, but they do not
give the correct $SU_2$ coupling $\tilde v_{SU_2}={k\over2}+16=14$.
As we have argued before,
the origin of the additional contribution $+16$ has to do with the fact
that the $SU_2$ that we actually observe is the diagonal subgroup
of the perturbative $SU_2$ and 16 non-perturbative $SU_2$'s
on the 7P's.

In the heterotic theory
the non-perturbative gauge fields do not contribute additional degrees of
freedom but they do show up in the value for $\tilde v$ of the
gauge factor with which they mix.
The gauge coupling constant of $SU_2^{\rm het}$ is thus
\eqn\gaugecoup{
{1\over g^2_{\rm het}}={1\over g^2_{10P}}+\sum_i {1\over g^2_{7P}}}
where the sum is over all those non-perturbative
gauge groups which mix with the perturbative gauge group
on the ten-plane. All coupling constants in eq.\gaugecoup\ are
in the six-dimensional theory. The subscripts on the right-hand side
refer to their origin.
For the model considered in this section, the sum is over
all sixteen non-perturbative $SU_2$ factors.
Also, ${1\over g^2_{10P}}={1\over\al'}
\Bigl({{\rm Vol(K3)}\over\lambda_H^2\al'^2}+\tilde v_{\rm pert}\Bigr)$
with $\tilde v_{\rm pert}={k\over2}$.
The seven-dimensional gauge couplings are ${1\over g^2|_{7P}}
\sim{R_{11}\over l_{11}^3}\sim{1\over\al'}$ where
$l_{11}$ is the eleven-dimensional Planck length.
We thus find
\eqn\vtilde{
\tilde v_{SU_2}={k_1\over 2}+\#(\hbox{7P groups that mix}).}
The non-perturbative contribution comes with an overall
coefficient one. This coefficient was fixed using supersymmetry, which
relates $\tilde v$ to the anomaly polynomial which gave
$\tilde v_{SU_2}=-2+16$. We give further supporting evidence for this
interpretation, based on the structure of the anomaly polynomial,
in Appendix~B.

It is now also straightforward to check that the above discussion
supports the distribution of fields that we have put forward.
\foot{We should however stress that we do not have a
truly M-theoretic derivation of these results.}
The non-perturbative gauge groups mix with the perturbative
$SU_2$ and thus must be locked to it on each I6 intersection plane.
For it not to produce extra massless degrees of freedom in the
heterotic limit $R_{11}\to 0$,
we must impose boundary conditions such that there
are no zero modes, neither for $V_7$ nor for $H_7$. Since, as discussed
above, they have opposite boundary conditions (free vs. fixed)
we must impose for both, the vector-multiplet and the hyper-multiplet,
Neumann conditions on one end of the $x^{11}$ interval and `Dirichlet'
on the other.\foot{Note that the locking condition \lock\ is not exactly
Dirichlet, but has the same effect on mode counting for the seven-dimensional
fields. Both, hyper and vector components, have half-integer modes only,
{\it i.e.} no zero-modes.}
In particular $V_7$ then has free boundary conditions on I6.
The situation for $H_7$ is reversed.

We have already pointed out one difference between the
compactification of the heterotic string on a smooth K3 and
on a singular K3. In the former case we always have
$\tilde v={k\over2}$ whereas this is not true in the latter case for those
gauge group factors which mix with the non-perturbative gauge
groups on the fixed seven-planes.
Another difference between the smooth and the singular geometry is
the fact that whereas in the former case the rank of the gauge group
is reduced, this is not so in the latter.
\foot{This presumes a $\IZ_N$ symmetric torus of generic size and
no $E_8\times E_8$ Wilson lines on $T^4$. In the completely generic case
of asymmetric orbifolds of a Narain $\Gamma_{4,20}$ compactification
of the heterotic string the resulting gauge group can have rank as
high as 20 or can be lower than 16.}
The rank reduction in the
smooth case is due to the presence of a non-trivial gauge bundle
with instanton numbers $n_{1,2}$. On the orbifold
a $\IZ_n$ singularity can support a singular $\IZ_n$
instanton with instanton number $1/n$,
which breaks the gauge group without reducing its
rank. {\it E.g.} for $\IZ_2$, the gauge group is  reduced to
$[E_7\times SU_2]/\IZ_2$ or ${\rm Spin}(16)/\IZ_2$, depending on the
choice of the $\IZ_2\subset E_8$ \BLPSSW.
{}For the model considered in this section the instanton number
on the $E_7\times SU_2$ side is eight, {\it i.e.} there is one
instanton of instanton number $1/2$ at each fixed point.
On the $SO_{16}$ side, we have instead one instanton of instanton
number one at each fixed point.

\newsec{Considerations of local anomaly cancellation}

One important consistency condition that the low-energy effective
field theory has to satisfy is anomaly freedom. This requirement is
particularly powerful in six dimensions since in addition to pure gauge
anomalies there are potential gravitational and mixed anomalies\AGW. They all
have to cancel. It is straightforward to verify that the anomaly of
the six-dimensional field theory factorizes and
the presence of one antisymmetric tensor in the massless spectrum
guarantees that this anomaly can be cancelled via the Green-Schwarz
mechanism.

In the M-theory description of the heterotic orbifold we have allocated all
massless fields (perturbative and non-perturbative)
to the bulk (gravity and moduli) and the
various types of planes which are present. We can now consider the
field theory on any one of these planes and  since anomalies
are a UV phenomenon, we need to require that they cancel locally, {\it i.e.}
on any plane separately. In the bulk and on the seven-planes this is
automatic, they are odd-dimensional. On each of the two ten-planes,
away from the intersection six-planes,
there are 16 supercharges and an entire $E_8$ gauge group.
Anomaly cancellation works in exactly the
same way as in the Ho\v{r}ava--Witten theory.
The situation on the intersection six-planes, however, involves
new features: here
supersymmetry is broken further to eight super-charges and the gauge group
is broken to a subgroup. The issue of anomaly cancellation on the six-planes
has to be addressed and in fact we will find that it provides a non-trivial
check on the scenario advocated in sect.~2.

We now turn to the evaluation of the anomaly on the intersection
six-planes.
It gets contributions from two sources. $(i)$ Quantum contributions:
they arise from the massless states which are charged under
the gauge group operating at the particular I6 or I6$'$ fixed plane we are
considering. Fields residing in the bulk, on the ten-plane into which the
six-plane is embedded, on the seven-plane which is bounded by the six-plane and
the fields confined to the six-plane do contribute.
$(ii)$ Inflow and intersection contributions:
they arise from gauge variance of the
11d SUGRA action. There is a contribution from a modified Bianchi identity
and contributions arising from Chern-Simons (CS) terms.
We first discuss the quantum contributions. Some basic results which we
will be using are collected in Appendix~A.

\noindent
(1) Bulk fields: gravity multiplet,
self-dual tensor multiplet and four moduli hyper-multiplets.
In this work we analyze only
compactifications with one tensor multiplet in the bulk, thus
restricting the discussion to the perturbative heterotic
string. Recall that in the Ho\v{r}ava--Witten theory the gravitational
anomaly is distributed evenly over the two
end-of-the-world ten-planes. By the same logic we
distribute the contribution of the bulk fields
between all $2\times 16$ fixed I6 and I6$'$ planes and obtain as
their contribution to the anomaly on each intersection six-plane\foot{The
correctly normalized anomaly polynomial in $d=6$ is
${-i\over192 \pi^3}{\cal A}$.}
\eqn\anone{\eqalign{
{\cal A}(\hbox{bulk})&={1\over2\cdot 16}
\left[-{244\over240}\tr R^4+{44\over192}(\tr R^2)^2\right]+
{4\over 2\cdot 16}\left[{1\over240}\tr R^4+{1\over 192}(\tr R^2)^2\right]\cr
&=-{1\over32}\tr R^4+{1\over128}(\tr R^2)^2\,.}}
The first term in the first line is the contribution from
the gravity multiplet and the tensor multiplet whereas the second term
is the contribution from the four moduli fields.

\noindent
(2) Ten-plane fields: these are  vector and hyper multiplets from the
untwisted sector.
We have to distinguish between I6 and I6$'$
planes, as they are embedded in 10P and 10P$'$, respectively,
which carry different perturbative gauge groups and untwisted  matter,
charged under $G$ (for I6) and $G'$ (for I6$'$).
${\cal N}=1$ supersymmetry demands that the fermions in a
vector-multiplet have opposite chirality than the fermions in a
hyper-multiplet and consequently they contribute
to the anomaly with opposite sign.
Thus generically
\eqn\generical{{\cal A}=(n_H-n_V)\left({1\over240}\tr R^4
+{1\over192}(\tr R^2)^2\right)
-{1\over4}\tr R^2\left(\Tr_H F^2-\Tr_V F^2\right)
+\left(\Tr_H F^4-\Tr_V F^4\right)}
where $n_H$ and $n_V$ is the number of hyper-multiplets and vector-multiplets,
respectively.
Distributing the anomaly of the untwisted fields charged under
$E_7\times SU_2$ evenly over the 16 I6 planes, we have
$H={1\over16}(\ul{56},\ul{2})$ and
$V={1\over16}[(\ul{133},\ul{1})+(\ul{1},\ul{3})]$.
Taking into account the gauge quantum numbers
we arrive, after a little algebra at\foot{See Appendix~C for the
group theory involved in this derivation.}
\eqn\antwoa{\eqalign{{\cal A}({\rm 10P})&=
-{1\over 160}\tr R^4-{1\over128}\big(\tr R^2\big)^2
+\tr R^2\Bigl({3\over32}\tr F_{\sss E_7}^2
-{13\over32}\tr F_{\sss SU_2}^2\Bigr)\cr
&\qquad-{3\over16}\bigl(\tr F_{\sss E_7}^2\bigr)^2
+{5\over16}\big(\tr F_{\sss SU_2}^2\big)^2
+{9\over8}\tr F_{\sss SU_2}^2\,\tr F_{\sss E_7}^2\,.}}
Similarly, at each I6$'$ plane with $G'=SO_{16}$ we have
$H={1\over16}(\ul{128})$ and $V={1\over16}(\ul{120})$, leading to
\eqn\antwob{
{\cal A}({\rm 10P}')={1\over480}\tr R^4+{1\over 384}\big(\tr R^2\big)^2
-{1\over32}\tr R^2\cdot\tr F_{\sss SO_{16}}^2
+{3\over16}\big(\tr F_{\sss SO_{16}}^2\big)^2-\tr F_{\sss SO_{16}}^4\,.}

\noindent
(3) Seven-plane fields:
each of the sixteen 7P's connects (across the $x^{11}$ direction)
an I6 plane to an I6$'$ plane.
As explained in the previous section, each 7P
carries an $SU^{\rm non-pert}_2$ with a
$d=7$ vector-multiplet whose reduction to
$d=6$ gives a vector-multiplet and an adjoint hyper-multiplet.
With the assignment of boundary conditions as specified before, the
hyper-multiplet component contributes to the anomaly on I6 whereas
the vector-multiplet component contributes on I6$'$. There is however
one subtlety concerning the precise contribution.
A priori the contribution from both the
vector and the hyper components is distributed evenly
over the two bounding six-planes.
The local boundary conditions however determine whether they do in
fact give rise to an anomaly. This is only the case if the fields satisfy
free boundary conditions. This is
a manifestation of the local consistency assumption:
what happens at a given boundary is sensitive only to the
boundary conditions imposed there and is blind to what happens at another,
distant boundary. In other words, the fact that each multiplet, due to the
chosen boundary conditions, contributes to the anomaly only on one of
the two ends does not affect the amount by which it contributes.
In summary, the contribution to the local anomaly from the 7P fields is
{\it half} of that of a $SU_2$ vector-multiplet on I6$'$ and 
{\it half} of that of an an $SU_2$ adjoint hyper-multiplet on I6.
We then find
\eqn\anthrees{
\hfilneg\eqalign{
\qquad\qquad{\cal A}(7P)&
=\hphantom{+}{1\over 160}\tr R^4+{1\over128}\big(\tr R^2\big)^2
-{1\over4}\tr R^2\cdot\tr F_{\sss SU_2}^2+\big(\tr F_{\sss SU_2}^2\big)^2\cr
\noalign{\smallskip\hbox{on the I6 and}\smallskip}
{\cal A}(7P)&
=-{1\over160}\tr R^4-{1\over128}\big(\tr R^2\big)^2
+{1\over4}\tr R^2\cdot\tr F_{\sss SU_2}^2-\big(\tr F_{\sss SU_2}\big)^2\cr
\noalign{\smallskip\hbox{on the I$6'$.}\smallskip}
}}

\noindent
(4) Six-plane fields:
these are the massless fields which are entirely confined to the I6 (I6$'$)
planes. In the models which we investigate there are no such tensor or vector
states and the only contribution arises from hyper-multiplets which appear
in the twisted sectors of the heterotic theory. For the
$\IZ_2$ model considered here the twisted sector contains
sixteen half-hyper-multiplets, one localized at each fixed point
and transforming as $(\ul{1},\ul{2};\ul{16})$. It should be assigned
to the I6$'$ side since it is charged under the gauge groups residing here,
namely ($SU_2\times SO_{16}$). There are no six-plane fields on I6.
Note that the $SU_2$ quantum numbers should be understood as those
pertaining to the diagonal $SU_2$ group. We thus find\foot{If there
were vector-multiplets confined to the six-planes, they
would contribute in an obvious way. If there were $n_{T_6}$
tensor multiplets, there would be an additional contribution of
$n_{T_6}\left({29\over 240}\tr R^4-{7\over 192}(\tr R^2)^2\right)$.}
\eqn\anfours{\eqalign{
{\cal A}(6P)&=0\cr
{\cal A}(6P')&={1\over15}\tr R^4+{1\over12}\big(\tr R^2\big)^2
-\tr R^2\Bigl(\tr F_{\sss SU_2}^2-{1\over4}\tr F_{\sss SO_{16}}^2\Bigr)
\qquad\qquad\cr
&\qquad+\tr F_{\sss SO_{16}}^4+\big(\tr F_{\sss SU_2}\big)^2
+{3\over2}\tr F_{\sss SU_2}^2\cdot \tr F_{\sss SO_{16}}^2\,.}}
This completes the enumeration of the quantum anomalies.

In addition to the quantum anomalies there are additional
contributions which have their origin in the gauge-variance of the classical
low-energy M-theory effective action \FOL.
More precisely, these `inflow' contributions arise from the
CS couplings $C\wedge G\wedge G$ and $C\wedge
(\tr R^4-{1\over4}(\tr R^2)^2)$ in the
eleven-dimensional bulk action.
Here $C$ is the three-form potential and $G$ its
field strength with (magnetic) sources on the ten-planes and the six-planes.
This produces an anomaly in the gauge theory on the six-planes \FOL
\eqn\anfives{
{\cal A}(\hbox{inflow})=-g\Bigl({1\over8}\tr R^4-{1\over 32}(\tr R^2)^2\Bigr)
-{3g\over4}\Bigl(\tr F_{10}^2-{1\over 2}\tr R^2\Bigr)^2\,.}
$F_{10}$ are the gauge fields on the ten planes only, {\it i.e.} $G$ or $G'$
and $g$ is the magnetic charge of the six-plane.
To the best of our
knowledge the coefficients of these terms have not been reliably determined.
We have fixed the normalizations
such as to make the anomaly cancellation work for this
model. Once fixed we use them for all other models which we consider.
\foot{We would like to point out that our relative normalization differs by
a crucial factor three from that of ref.~\FOL.}
Below we will give independent arguments in support of the 
normalizations in \anfives.
The magnetic charges are easy to determine. We have found in sect.~2.2 that
the magnetic charge of the ten-planes is $k_{1,2}$.
In the orbifold limit the geometric and the gauge curvatures are
restricted to the orbifold singularities. Hence this is also where the
magnetic charge is sitting. We thus get the following sum rule for the
magnetic charges of the six-planes
\eqn\gks{
\sum_{i} g_{i\al}=k_\al\,.}
The sum extends over all six-planes on a given side of the $x^{11}$ interval
and $\al=1,2$.
{}For the problem at hand $k_{1,2}=\pm 4$ and since all I6 are related
by symmetry (and likewise the I6$'$) we find
$g_{\rm I6}={1\over16}\cdot(-4)=-{1\over 4}$ and
$g_{\rm{I6}'}={1\over16}\cdot 4=+{1\over4}$. We thus obtain for the
inflow contributions to the anomalies on the six-planes:
\eqn\inflows{
\hfilneg\!\eqalign{
{\cal A}({\rm inflow})&
=\hphantom{+}{1\over32}\tr R^4-{1\over128}\big(\tr R^2\big)^2
+{3\over16}\Bigl(\tr F_{\sss E_7}^2+\tr F_{\sss SU_2}^2
-{1\over2}\tr R^2\Bigr)^2\cr
\omit on each I6 plane and\quad\hfil\cr
{\cal A}({\rm inflow})&
=-{1\over32}\tr R^4+{1\over128}\big(\tr R^2\big)^2
-{3\over16}\Bigl(\tr F_{\sss SO_{16}^2}-{1\over2}\tr R^2\Bigr)^2\cr
\omit on each I$6'$ plane.\quad\hfil\cr
}\hskip -1em
}
Adding the quantum and the inflow anomalies, we obtain
\eqn\qplusin{
\hfilneg {\rm and}\quad\hfil
\eqalign{
{\cal A}(\rm{quantum+inflow})&
={3\over2}\Bigl(\tr F_{\sss SU_2}^2
-{1\over16}\tr R^2\Bigr)\cdot\Bigl(\tr F_{\sss E_7}^2+\tr F_{\sss SU_2}^2
-{1\over2}\tr R^2\Bigr)\cr
\noalign{\vskip\baselineskip}
{\cal A}(\rm{quantum+inflow})&
={3\over2}\Bigl(\tr F_{\sss SU_2}^2
-{1\over16}\tr R^2\Bigr)\cdot\Bigl(\tr F_{\sss SO_{16}}^2
-{1\over2}\tr R^2\Bigr)\,.\cr
}}
Note that these expressions factorize into two terms, the first of
a seven-plane origin and the second of a ten-plane origin.

Consistency of the theory requires, in the absence of tensor multiplets
on the I6 and I6$'$ planes, that the anomaly vanishes identically.\foot{In the
presence of extra tensor multiplets one could invoke the GS mechanism
provided the anomaly factorizes appropriately.}
Thus, were it only for these contributions to the local anomaly, the theory
would be inconsistent. The factorization pattern of \qplusin\
does suggest yet another contribution
to the anomaly. Indeed, we still have to take into account the
intersection anomaly \FOL.
It arises from the electric coupling
of the 7P to $C$. This gives rise to a CS term on each 7P world-volume
of the form $C\wedge Y_4$ which leads to a contribution to the anomaly
on the intersection six-planes
\eqn\ansixs{{\cal A}(\hbox{intersection})=\Bigl(\tr F_{10}^2
-{1\over2}\tr R^2\Bigr)\times Y_4}
with
\eqn\ysevens{
Y_4={3\over2}(\eta\,\tr R^2-\rho\,\tr F_7^2)\,.}
Here $F_7$ refers to the 7P gauge fields and $F_{10}$ to the
ten-plane gauge fields.
$\eta$ and $\rho$ are as yet free parameters. They should follow
from seven-dimensional physics and should in particular not depend on the
details of the boundary conditions imposed at the ends of the interval.
We will come back to this issue in sect.~4.
However, comparing eqs.\qplusin\ and \ansixs\ requires that for the
$\IZ_2$ orbifold model considered here,
\eqn\cancellations{
\eta={1\over16}\quad\hbox{and}\quad
\rho=1\,.}

So far we have presented a consistent scenario for the $M$-theory
description of the heterotic $T^4/\IZ_2$ orbifold with gauge
group $[E_7\times SU_2]\times SO_{16}$. In the remaining sections
we will generalize the analysis to other $T^4/\IZ_N$ orbifolds.

\newsec{Examples of $\IZ_3,\,\IZ_4$ and $\IZ_6$ orbifolds}

In this section we generalize the discussion of the two preceding sections
to orbifolds with higher order fixed points.
In 4.1. we recall some basic facts about $T^4/\IZ_N$ orbifolds. We then
discuss one model for $N=3,4,6$ each.
We will stress the new issues which
arise for non-prime orbifolds, {\it i.e.} for $N=4,6$.
We first analyse the gauge
couplings (sect.~4.2) and then turn to the analysis of local anomalies
(sect.~4.3) of these models.
In sect.~4.4 we `derive' the values for the
parameters $\rho$ and $\eta$ in \ysevens. We give two independent arguments.
The first is based on a comparison between the heterotic anomaly and
the M-theory anomaly. The second argument does not rely on anomaly
considerations but uses the duality between
M-theory on K3 and heterotic theory on $T^3$.
To help the reader through the discussion of the various models, we have
collected the data concerning their spectra, gauge couplings and
magnetic charges in three tables. They can be found in sect.~4.2.

\subsec{Some facts about $T^4/\IZ_N$ orbifolds}

To ensure $d=6$, ${\cal N}=1$ supersymmetry (eight unbroken
supercharges) in $T^4/\IZ_N$ compactifications of the heterotic string,
$N$ has to be  restricted to  $N\in\lbrace 2,3,4,6\rbrace$.
Whereas for the $\IZ_2$ orbifold the number of moduli is four, for
the remaining cases there are only two moduli.
This follows by considering which of the (1,1)
forms on the torus are invariant under the
$\IZ_N$ twist. Recall that if $T^4$ is parametrized by two
complex coordinates $z^1$ and $z^2$, the discrete $\IZ_N$ transformation
acts as $z^1\to \al z^1,\, z^2\to \al^{N-1}z^2$,
where $\al^N=1$. The action is such that $dz^1\wedge dz^2$
survives, since this is the (2,0) form which we need for
a K3 compactification or the orbifold limits thereof. Its presence
also guarantees at least eight unbroken supercharges.
There are various ways to embed the geometrical twists as shifts in the
gauge lattice, leading to different spectra. In the absence of
Wilson lines there are 2 (5,12,59)
different embeddings for $\IZ_2$ ($\IZ_3,\,\IZ_4,\,\IZ_6$)
with different gauge groups $G\times G'\subset E_8\times E_8$
and matter content \AFIUV,\stieberger.
The number of fixed points is given by the Lefshetz fixed-point formula
as
\smallskip
\eqn\lefschetz{
(1-\al)^2(1-\al^*)^2
=\cases{ 4^2 & for $\IZ_2$\cr
         3^2 & for $\IZ_3$\cr
         2^2 & for $\IZ_4$\cr
         1 & for $\IZ_6$\cr}}
\smallskip\noindent
For the other possible
lattice automorphisms, $N\in\lbrace 5,8,10,12\rbrace$,
the fixed point formula \lefschetz\ does not give an integer.
The precise fixed-point structure for $N$ non-prime will be discussed
when we consider the $\IZ_4$ and $\IZ_6$ examples below.

Much of the M-theory discussion from sect.~2 carries over verbatim to the
general case. The fixed seven-planes are again denoted by 7P and
they carry a non-perturbative gauge group, $G_7$,
which for a $\IZ_N$ fixed point is (at most) $SU_N$.
The intersection six-planes are again
I6 (at $x^{11}=0$) and I6$'$ (at $x^{11}=\pi R_{11}$).
There is one I6, I6$'$ pair for each orbifold fixed point.
For non-prime orbifolds there are fixed points of different orders. This
leads to new features which we will discuss in detail below. But first we
present another prime orbifold.

\subsec{Analysis of the gauge couplings}

\noindent
{\it $\IZ_3$ orbifold:}

The discussion here follows very closely
the discussion of the $\IZ_2$ orbifold in sect.~2. We will
thus be brief.
Since this is a prime orbifold, there is only one type
of fixed points, namely nine $\IZ_3$ fixed points.
Consequently there are nine seven-planes. They are bounded by nine
I6 planes and nine
I6$'$ planes. The I6 are related by symmetry
and are thus completely equivalent and likewise for the I6$'$.
Two of the four moduli of the
$T^4$ are invariant under the $\IZ_3$ twist. If we choose
$\delta={1\over3}(-2,1,1,0,0,0,0,0;
{5\over2},{1\over2},{1\over2},{1\over2},{1\over2},
{1\over2},{1\over2},{1\over2})$ as shift vector, the gauge group is
$G\times G'=(E_6\times SU_3)\times SU_9$. In addition to the states in the
corresponding vector-multiplets, there are twist-invariant massless states
which comprise hyper-multiplets transforming as
$(\ul{27},\ul{3};\ul{1})$ and $(\ul{1},\ul{1};\ul{84})$.
As for the massless twisted matter there is one hyper-multiplet
$(\ul{1},\ul{3};\ul{9})$ located at each of the nine fixed points.
This part of the spectrum and other data pertaining to this
model are summarized in table 4.1. below.

Given the massless spectrum, it is straightforward to compute the
one-loop beta-function coefficients for each group factor
and from this, via \bv, the $\tilde v_\al$, c.f. \gaugekin. We find
\eqn\besZthree{\eqalign{
b(E_6)=-3\quad & \rightarrow \quad\tilde v(E_6)=-{3\over 2}\,,\cr
b(SU_3)=51\quad & \rightarrow \quad\tilde v(SU_3)=-{3\over2}+9\,,\cr
b(SU_9)=15\quad & \rightarrow \quad\tilde v(SU_9)=+{3\over2}\,.}}
We thus learn that $\tilde v_{E_7}={k_1\over2}$ and
$\tilde v_{SU_9}={k_2\over 2}$ with $k_1=-3$ and $k_2=+3$,
but $\tilde v_{SU_3}={k_1\over2}+9$. In analogy
with the $\IZ_2$ example we conclude that the heterotic $SU_3$ gauge factor
is a linear combination of the perturbative $SU_3$ and the nine
non-perturbative $SU_3$'s on the seven-planes; {\it i.e.}
$(SU_3)_{\rm het.}
={\rm diag}[SU_3^{\rm pert}\times (SU_3^{\rm non-pert})^9]$.
Since the twisted hyper-multiplets are charged under $SU_9$, they
must be located on the I6$'$ planes and they carry quantum numbers
of a non-perturbative $SU_3$.
The boundary conditions of the hyper and the vector
components under the $7d\to 6d$ decomposition must be chosen such that
the vector component enjoys Dirichlet
boundary conditions of the type \lock\ on the $E_6\times SU_3$
side and Neumann conditions on the $SU_9$ side; {\it i.e.} there is one
$SU_3$ adjoint hyper-multiplet on each I6 and one vector-multiplet
on each I6$'$. This allocation of fields leads to
local anomaly cancellation on the six-planes as we will demonstrate
in sect.~4.3.

%
%
$$
\vbox{\offinterlineskip\tabskip=0pt
\halign{\strut
\vrule width0.8pt \quad #
\tabskip=0pt plus 100pt
& \unskip\quad $#$ \hfill \quad
&\vrule#& \quad \hfill $#$ \hfill\quad
&\vrule#& \quad \hfill $#$ \hfill\quad
&\vrule width0.8pt#\cr
\noalign{\hrule}\noalign{\hrule}
& \multispan 5 \hfill \hbox{Example 2:~~} $\IZ_3$ \hbox{orbifold}\hfill &\cr
\noalign{\hrule}\noalign{\hrule}
& \hbox{shift vector} && (-{2\over3},{1\over3},{1\over3},0,0,0,0,0) &&
 ({5\over6},{1\over6},{1\over6},{1\over6},{1\over6},{1\over6},
{1\over6},{1\over6})&\cr
\noalign{\hrule}
& \hbox{gauge group} && \hfill E_6\times SU_3 \hfill &&
\hfill SU_9\hfill &\cr
\noalign{\hrule}
& \hbox{matter} \smash{\lower7pt
\hbox{$~~~H_0\quad\cases{\phantom{.}&\cr\phantom{.}&\cr
\phantom{.}&\cr}$}} \!\!\!\!\!\!\!\!\!\!\!\!\!\!\!\!
&& \multispan 3 \hfill\hbox{two moduli} \hfill &\cr
& &&
(\ul{27},\ul{3};\ul{1}) & \omit & (\ul{1},\ul{1};\ul{84}) &\cr
& \hfill ~~~~~~~~~H_1 &&
\multispan 3 \hfill\hbox{$9\times(\ul{1},\ul{3};\ul{9})$}\hfill&\cr
\noalign{\hrule}
& b(G) && b(E_6)=-3,\,\, b(SU_3)=51 &&  b(SU_9)=15 &\cr
& \tilde v^{~}_G && \tilde v^{~}_{E_6}=-{3\over2},\,
\tilde v^{~}_{SU_3}=-{3\over2}+9
&& \tilde v^{~}_{SU_9}=+{3\over2} &\cr
& k=n-12 && -3 && +3 &\cr
& g^{~}_{\rm I6} && -{1\over 3} && +{1\over 3} &\cr
\noalign{\hrule}
&Q_{10} && {1\over 9}\lbrace(\ul{27},\ul{3})
-(\ul{78},\ul{1})-(\ul{1},\ul{8})\rbrace &&
{1\over9}\lbrace\cdot(\ul{84})-(\ul{80})\rbrace
&\cr
&Q_6 && \hbox{---}
&& (\ul{1},\ul{3};\ul{9})&\cr
\noalign{\hrule}
& G_7 && \multispan 3 \hfill\hbox{$SU_3$}\hfill &\cr
\noalign{\hrule}
&Q_7 && {1\over2}\cdot\ul{8}&& -{1\over 2}\cdot\ul{8}&\cr
\noalign{\hrule}
\noalign{\hrule}
}
\vskip.2cm
\noindent
{\ninepoint {\bf Table 4.1}: The $\IZ_3$ orbifold with gauge group
$(E_6\times SU_3)\times SU_9$.
The magnetic charges\hfill\break
will be computed in sect.~4.3, where also the
notation $Q_{10}$ etc. will be explained.}
}
$$

\medskip
\noindent
{\it $\IZ_4$ orbifold}

This is our first example of a non-prime orbifold. It possesses
two types of fixed points: four $\IZ_4$ fixed points and
16 $\IZ_2\subset\IZ_4$ fixed-points. The latter are obviously also
fixed under $\IZ_2$ and the twelve remaining $\IZ_2$ fixed points
lie pair-wise on $\IZ_4$ orbits.
There are thus four $\IZ_4$ and six $\IZ_2$ fixed points. All
fixed points of a given type are completely equivalent.
The two type we have to treat separately, though. The results of the
ensuing below of a particular $\IZ_4$ orbifold model
are collected in table 4.2 below.

We embed the $\IZ_4$ twist into the gauge sector via the shift vector
$\delta={1\over4}(-3,1,1,1,0,0,0,0;$ $-{7\over2},{1\over2},{1\over2},{1\over2},
{1\over2},{1\over2},{1\over2},{1\over2})$, which leads to
a breaking of the gauge group
$E_8\times E_8\to (SO_{10}\times SO_6)\times(SU_8\times SU_2)=G\times G'$.
The massless charged matter in the untwisted sector consists of
one hyper-multiplet transforming as $(\ul{16},\ul{4};\ul{1},\ul{1})$
and one hyper-multiplet transforming as $(\ul{1},\ul{1};\ul{28},\ul{2})$.
These states, which we denote by $H_0$, clearly live on 10P$'$.

Since for non-prime orbifolds there are different types of fixed-points,
one has to be careful with the allocation of the twisted matter.
The $\IZ_4$ orbifold has three
types of twisted sectors. States in the first and the third twisted
sectors combine into particle and anti-particle pairs and into
complete hyper-multiplets, which we denote as $H_1$.
These states are necessarily located at the $\IZ_4$ fixed points.
Straightforward application of the rules reviewed in \AFIQ\
gives $H_1=4\times(\ul{1},\ul{4};\ul{8},\ul{1})$.
The double-twisted sector contains both
particles and their anti-particles. These states we call $H_2$.
One finds $H_2=10\times{1\over2}(\ul{1},\ul{6};\ul{1},\ul{2})
+6\times{1\over2}(\ul{10},\ul{1};\ul{1},\ul{2})$.
The correct assignment of these states to the various fixed
points is as follows. At each $\IZ_2$ fixed point:
${1\over2}(\ul{1},\ul{6};\ul{1},\ul{2})
+{1\over2}(\ul{10},\ul{1};\ul{1},\ul{2})$ and
at each $\IZ_4$ fixed point: ${1\over2}(\ul{1},\ul{6};\ul{1},\ul{2})$.

Note that the states at each of the $\IZ_2$ fixed points
combine into representations of $SO_{16}\times(E_7\times SU_2)$ which
is the group left unbroken by the shift $2\delta$:
${1\over2}[(\ul{1},\ul{6};\ul{1},\ul{2})
+(\ul{10},\ul{1};\ul{1},\ul{2})]_{(SO_{10}\times SO_6)\times(SU_8\times SU_2)}
={1\over2}(\ul{16};\ul{1},\ul{2})_{SO_{16}\times(E_7\times SU_2)}$.
Also, locally at the $\IZ_2$ fixed points, there are four moduli
hyper-multiplets. In fact, the local physics at each of
the six $\IZ_2$ fixed points of this $\IZ_4$ orbifold is identical to
that encountered in the $\IZ_2$ model discussed before.
Since there are six $\IZ_2$ seven-planes, we expect
$\tilde v_{SU_2}={k_2\over2}+6$. We will verify this shortly.

It remains to discuss the four $\IZ_4$ fixed points
with twisted massless matter
$(\ul{1},\ul{4};\ul{8},\ul{1})+{1\over2}(\ul{1},\ul{6};\ul{1},\ul{2})$.
Each of the $\IZ_4$ seven-planes supports a non-perturbative
$SU_4\sim SO_6$
gauge group. This, together with the gauge quantum numbers of the
twisted states, suggests that the twisted matter lives on the
I6$'$ planes and transforms as $\ul{4}$ and $\ul{6}$ under a
non-perturbative $SU_4$, respectively. Also, in complete analogy to the
previous examples, this would imply
$\tilde v_{SU_4}={k_1\over2}+4$ and thus
$SU_4^{\rm het}={\rm diag}[SU_4^{\rm pert}
\times(SU_4^{\rm non-pert})^4]$.
That this is correct can be easily
verified, given that
\eqn\besZfour{\eqalign{
b(SO_{10})=6\quad & \rightarrow \quad\tilde v(SO_{10})=0\,,\cr
b(SO_6)=30\quad & \rightarrow \quad\tilde v(SO_6)=0+4\,,\cr
b(SU_8)=6\quad & \rightarrow \quad\tilde v(SU_8)=0\,,\cr
b(SU_2)=42\quad & \rightarrow \quad\tilde v(SU_2)=0+6\,.}}
We also find $k_1=k_2=0$, {\it i.e.} the 24 instantons are distributed
evenly over the two sides. It is now also clear how to distribute the
seven-plane fields. A $SU_4$ adjoint hyper-multiplet lives on
each of the $\IZ_4$ I6 planes and a $SU_4$ vector-multiplet on each
$\IZ_4$ I6$'$ plane.

%
%
$$
\vbox{\offinterlineskip\tabskip=0pt
\halign{\strut
\vrule width0.8pt \quad #
\tabskip=0pt plus 100pt
& \unskip\quad $#$ \hfill \quad
&\vrule#&\quad\hfill $#$ \hfill\quad
&\vrule#& \quad \hfill $#$ \hfill\quad
&\vrule width0.8pt#\cr
\noalign{\hrule}\noalign{\hrule}
& \multispan 5 \hfill \hbox{Example 3:~~} $\IZ_4$ \hbox{orbifold}\hfill &\cr
\noalign{\hrule}\noalign{\hrule}
& \hbox{shift vector} && (-{3\over4},{1\over4},{1\over4},{1\over4},0,0,0,0) &&
 (-{7\over8},{1\over8},{1\over8},{1\over8},{1\over8},{1\over8},
{1\over8},{1\over8})&\cr
\noalign{\hrule}
& \hbox{gauge group} && \hfill SO_{10}\times SU_4\hfill &&
\hfill SU_8\times SU_2\hfill &\cr
\noalign{\hrule}
& \hbox{matter} \smash{\lower7pt
\hbox{$~~~H_0\quad\cases{\phantom{.}&\cr\phantom{.}&\cr
\phantom{.}&\cr}$}} \!\!\!\!\!\!\!\!\!\!\!\!\!\!\!\!
&& \multispan 3 \hfill\hbox{two moduli} \hfill &\cr
& &&
(\ul{16},\ul{4};\ul{1},\ul{1}) & \omit & (\ul{1},\ul{1};\ul{28},\ul{2}) &\cr
& \hfill ~~~~~~~~~H_1 &&
\multispan 3 \hfill\hbox{$4\times(\ul{1},\ul{4};\ul{8},\ul{1})$}\hfill&\cr
& \hfill ~~~~~~~~~H_2 &&
\multispan 3 \hfill\hbox{$10\times{1\over2}(\ul{1},\ul{6};\ul{1},\ul{2})
+6\times{1\over2}(\ul{10},\ul{1};\ul{1},\ul{2})$}\hfill&\cr
\noalign{\hrule}
& b(G) && b(SO_{10})=6,\,\, b(SU_4)=30 &&  b(SU_8)=6,\,\, b(SU_2)=42 &\cr
& \tilde v^{~}_G && \tilde v^{~}_{SO_{10}}=0,\,
\tilde v^{~}_{SU_4}=0+4
&& \tilde v^{~}_{SU_8}=0,\,\tilde v^{~}_{SU_2}=0+6 &\cr
& k=n-12 && 0 && 0 &\cr
\noalign{\hrule}
& \multispan 5 \hfill \hbox{$\IZ_2$} \hbox{fixed points:}~~
\hbox{see Sects. 2 and 3}\hfill&\cr
\noalign{\hrule}
& \multispan 5 \hfill \hbox{$\IZ_4$} \hbox{fixed points:}\hfill&\cr
\noalign{\hrule}
& g^{~}_{\rm I6} && -{3\over 8} && +{3\over 8} &\cr
\noalign{\hrule}
&Q_{10} && -{5\over 32}\lbrace(\ul{45},\ul{1})
+(\ul{1},\ul{15})\rbrace && -{5\over32}\lbrace(\ul{63},\ul{1})
+(\ul{1},\ul{3})\rbrace &\cr
& &&+{3\over32}(\ul{10},\ul{6})
+{1\over16}(\ul{16},\ul{4})
&&
+{3\over32}(\ul{70},\ul{1})+{1\over16}(\ul{28},\ul{2})
&\cr
&Q_6 && \hbox{---}
&& (\ul{1},\ul{4};\ul{8},\ul{1})+{1\over2}(\ul{1},\ul{6};\ul{1},\ul{2})&\cr
\noalign{\hrule}
& G_7 && \multispan 3 \hfill\hbox{$SU_4$}\hfill &\cr
\noalign{\hrule}
&Q_7 && {1\over2}\cdot\ul{15}&& -{1\over 2}\cdot\ul{15}&\cr
\noalign{\hrule}
\noalign{\hrule}
}
\vskip.2cm
\noindent
{\ninepoint {\bf Table 4.2}: The $\IZ_4$ orbifold with gauge group
$(SO_{10}\times SU_4)\times (SU_8\times SU_2)$.}
}
$$

\medskip
\noindent
{\it $\IZ_6$ orbifold}

As in the previous examples, the data of the particular
model we will discuss in this section are collected in a table
which can be found at the end of this subsection.

As $\IZ_6$ has two non-trivial subgroups, $\IZ_2$ and $\IZ_3$,
a $\IZ_6$ orbifold has fixed points of orders 2, 3 and 6. The
Lefshetz fixed point theorem gives one $\IZ_6$ fixed point
which is of course also fixed under the $\IZ_2$ and $\IZ_3$ subgroups.
The remaining eight $\IZ_3$ fixed points lie on four $\IZ_6$ orbits.
The 15 $\IZ_2$ fixed points not fixed under $\IZ_6$ lie on 5 $\IZ_6$ orbits.
A $\IZ_6$ orbifold thus has one $\IZ_6$, four $\IZ_3$ and five $\IZ_2$
fixed points. Of the four moduli of $T^4$, two are invariant under
the $\IZ_6$ twist, {\it i.e.} we have two moduli hyper-multiplets.

To proceed, we need to specify the shift vector. Our choice
$\delta={1\over6}(-5,1,1,1,1,1,0,0;$ $-5,1,1,1,1,1,1,1)$
leads to the gauge group
$G\times G'=(SU_6\times SU_3\times SU_2)\times SU_9$.
Locally at the $\IZ_3$ fixed points the gauge group is that corresponding to
the shift vector $2\delta$. One finds $(E_6\times SU_3)\times SU_9$
and thus recovers the situation of the $\IZ_3$ orbifold model
discussed above. At the $\IZ_2$ fixed points the shift $3\delta$ leads to
the gauge group $(E_7\times SU_2)\times E_8$, {\it i.e.}
the $\IZ_2\subset\IZ_6$ subgroup leaves the second $E_8$ unbroken.
We will discuss this
$\IZ_2$ orbifold in some detail in sect.~5. For the present
purposes it suffices to state the following facts.
The untwisted matter for this $\IZ_2$ model
is $H_0=(\ul{56},\ul{2};\ul{1})$ and the twisted matter is
$H_1=16\times\lbrace{1\over2}(\ul{56},\ul{1};\ul{1})
+2(\ul{1},\ul{2};\ul{1})\rbrace$. Since the second $E_8$ is unbroken,
all 24 instantons must sit in the first $E_8$, {\it i.e.} $k=12$.
Finally, the $SU_2$ gauge-factor is purely perturbative,
{\it i.e.} it does not mix with the seven-plane gauge group.

The untwisted massless matter of the $\IZ_6$ orbifold consists of
a single hyper-multiplet $H_0=(\ul{6},\ul{3},\ul{2};\ul{1})$
which lives on 10P. The twisted matter states are
$H_1=(\ul{6},\ul{1},\ul{1};\ul{\bar 9})$: this hyper-multiplet is
necessarily located at the $\IZ_6$ fixed point;
$H_2=4\times(\ul{1},\ul{3},\ul{1};\ul{9})$: there is one such hyper-multiplet
at each of the four $\IZ_4$ fixed points;
the states in the third-twisted sector,
$H_3=6\times{1\over2}(\ul{20},\ul{1},\ul{1};\ul{1})
+5\times (\ul{6},\ul{3},\ul{1};\ul{1})+10(\ul{1},\ul{1},\ul{2};\ul{1})$,
are assigned to the different fixed points as follows. At each of the five
$\IZ_2$ fixed points there is a ${1\over2}(\ul{56},\ul{1})
+2(\ul{1},\ul{2})$ hyper-multiplet of $E_7\times SU_2$.
Under $E_7\times SU_2\to SU_6\times SU_3\times SU_2$ it
decomposes as
${1\over2}(\ul{20},\ul{1},\ul{1})
+(\ul{6},\ul{3},\ul{1})+2(\ul{1},\ul{1},\ul{2})$. This leaves one
${1\over2}(\ul{20},\ul{1},\ul{1};\ul{1})$ half-hyper-multiplet at
the $\IZ_6$ fixed points. To summarize, the massless twisted
matter content at the $\IZ_6$ fixed point is
$(\ul{6},\ul{1},\ul{1};\ul{\bar 9})
+{1\over2}(\ul{20},\ul{1},\ul{1};\ul{1})$.

The coefficients of the one-loop beta-functions and the
resulting values of $\tilde v_\al$ are easily found to be
\eqn\besZfour{\eqalign{
b(SU_{6})=18\quad & \rightarrow \quad\tilde v(SU_{6})=1+1\,,\cr
b(SU_3)=36\quad & \rightarrow \quad\tilde v(SU_3)=1+4\,,\cr
b(SU_2)=12\quad & \rightarrow \quad\tilde v(SU_2)=1\,,\cr
b(SU_9)=0\quad & \rightarrow \quad\tilde v(SU_9)=-1\,.}}
Since the maximal non-perturbative gauge group is $SU_6$,
we conclude that the value of $\tilde v_{SU_9}$ must be that
of a perturbative $SU_9$ factor. Consequently
$\tilde v_{SU_9}={k_2\over2}$ and $k_2=-2,\,k_1=+2$. From $\tilde v_{SU_6}=2$
it then follows that the non-perturbative $SU_6$ mixes with the perturbative
$SU_6$, {\it i.e.}
$SU_6^{\rm het}={\rm diag}[SU_6^{\rm pert}\times SU_6^{\rm non-pert}]$.
The fact that $\tilde v_{SU_3}=1+4$ is in agreement with our expectation
that four non-perturbative $SU_3$ factors located at the four $\IZ_3$
seven-planes mix with the perturbative $SU_3$.
As for $\tilde v_{SU_2}$ we would naively expect
$\tilde v_{SU_2}={k_1\over2}+5$. This would in fact be required if at the
$\IZ_2$ fixed points the local physics were that of the $\IZ_2$ orbifold
discussed in sect.~2.
However, as already mentioned above, for the $\IZ_2$ orbifold with
gauge group $(E_7\times SU_2)\times E_8$ we will find
in sect.~5 that the $SU_2$ factor is completely perturbative, {\it i.e.}
there is no mixing with the non-perturbative seven-plane gauge group.

It is now straightforward to give the field content on the
$\IZ_6$ I6 and I6$'$ planes. On I6 with perturbative gauge group
$SU_6\times SU_3\times SU_2$ there are no twisted matter states.
They are all located at I6$'$. Since they carry $SU_6$ quantum
numbers, the non-perturbative $SU_6$ vector-multiplet must be free
on I6$'$ and the $SU_6$
adjoint hyper-multiplet is free on I6.
%
%
$$
\hskip 0pt minus 0.5in
\vbox{\offinterlineskip\tabskip=0pt
\halign{\strut
\vrule width0.8pt \quad #
\tabskip=0pt plus 100pt
& \unskip\quad $#$ \hfill \quad
&\vrule#&\hfill $#$ \hfill
&\vrule#& \quad \hfill $#$ \hfill\quad
&\vrule width0.8pt#\cr
\noalign{\hrule}\noalign{\hrule}
& \multispan 5 \hfill \hbox{Example 4:~~} $\IZ_6$ \hbox{orbifold}\hfill &\cr
\noalign{\hrule}\noalign{\hrule}
& \hbox{shift vector} && (-{5\over6},{1\over6},{1\over6},{1\over6},
{1\over6},{1\over6},0,0) &&
 (-{5\over6},{1\over6},{1\over6},{1\over6},{1\over6},{1\over6},
{1\over6},{1\over6})&\cr
\noalign{\hrule}
& \hbox{gauge group} && \hfill SU_6\times SU_3\times SU_2\hfill &&
\hfill SU_9\hfill &\cr
\noalign{\hrule}
& \hbox{matter} \smash{\lower7pt
\hbox{$~~~H_0\quad\cases{\phantom{.}&\cr\phantom{.}&\cr
\phantom{.}&\cr}$}} \!\!\!\!\!\!\!\!\!\!\!\!\!\!\!\!
&& \multispan 3 \hfill\hbox{two moduli} \hfill &\cr
& &&
(\ul{6},\ul{3},\ul{2};\ul{1}) & \omit & &\cr
& \hfill ~~~~~~~~~H_1 &&
\multispan 3 \hfill~~~~~~~~~~
\hbox{$(\ul{6},\ul{1},\ul{1};\ul{\bar 9})$}\hfill&\cr
& \hfill ~~~~~~~~~H_2 &&
\multispan 3 \hfill~~~~~~~~~~
\hbox{$4(\ul{1},\ul{3},\ul{1};\ul{9})$}\hfill&\cr
& \hfill ~~~~~~~~~H_3 &&
\hfill\hbox{$6\times{1\over2}(\ul{20},\ul{1},\ul{1};\ul{1})
+5(\ul{6},\ul{3},\ul{1};\ul{1})$}\hfill &\omit & &\cr
& && \hfill +10(\ul{1},\ul{1},\ul{2};\ul{1})\hfill &\omit& &\cr
\noalign{\hrule}
& b(G) &&\,\, b(SU_{6})=18,\,\, b(SU_3)=36,\,\,
b(SU_2)=12 &&  b(SU_9)=0 &\cr
& \tilde v^{~}_G && \tilde v^{~}_{SU_{6}}=1+1,\,
\tilde v^{~}_{SU_3}=1+4,\,\tilde v^{~}_{SU_2}=1
&& \tilde v^{~}_{SU_9}=-1 &\cr
& k=n-12 && 2 && -2 &\cr
\noalign{\hrule}
& \multispan 5 \hfill \hbox{$\IZ_2$} \hbox{fixed points:}~~
\hbox{see sect.~5}\hfill&\cr
\noalign{\hrule}
& \multispan 5 \hfill \hbox{$\IZ_3$} \hbox{fixed points:}~~
\hbox{see Example 1}\hfill&\cr
\noalign{\hrule}
& \multispan 5 \hfill \hbox{$\IZ_6$} \hbox{fixed point:}\hfill&\cr
\noalign{\hrule}
& g^{~}_{\rm I6} && -{5\over 12} && +{5\over 12} &\cr
\noalign{\hrule}
&Q_{10} && -{35\over 144}\lbrace(\ul{35},\ul{1},\ul{1})
+(\ul{1},\ul{8},\ul{1})+(\ul{1},\ul{1},\ul{3})\rbrace
&& {13\over72}(\ul{84})-{35\over144}(\ul{80})&\cr
& &&-{5\over72}(\ul{6},\ul{3},\ul{2})+{13\over72}(\ul{15},\ul{\bar 3},\ul{1})
+{19\over144}(\ul{20},\ul{1},\ul{2})
&& &\cr
&Q_6 && \hbox{---}
&& (\ul{6},\ul{1},\ul{1};\ul{\bar{9}})
+{1\over2}(\ul{20},\ul{1},\ul{1};\ul{1})&\cr
\noalign{\hrule}
& G_7 && \multispan 3 \hfill\hbox{$SU_6$}\hfill &\cr
\noalign{\hrule}
&Q_7 && {1\over2}\cdot\ul{35}&& -{1\over 2}\cdot\ul{35}&\cr
\noalign{\hrule}
\noalign{\hrule}
}
\vskip.2cm
\noindent
{\ninepoint {\bf Table 4.3}: The $\IZ_6$ orbifold with gauge group
$(SU_6\times SU_3\times SU_2)\times SU_9$.}
}\hskip 0pt minus 0.5in
$$

\subsec{Local anomaly cancellation}

We will now generalize the discussion of sect.~3 to the
models of the previous subsection.
In particular we will confirm the relative normalization
of the two contributions in \anfives.

\noindent
{\it $\IZ_3$ orbifold}

The discussion of the local anomalies for this model is almost
identical to the one given in sect.~3. One difference is
that now we have only two moduli multiplets. Also, when distributing
bulk fields and ten-plane fields over the various six-planes we have
to take into account that we now have nine fixed points.
With $k_1=-3$ and $k_2=+3$, the magnetic charges of the
nine I6 and the nine I6$'$ planes are $g_{\rm I6}=-{1\over3}$
and $g_{{\rm I6}'}=+{1\over3}$, respectively.

It is straightforward to determine the quantum + inflow
contribution to the local anomaly on the six-planes to
\eqn\qplusinzt{
\eqalign{
{\cal A}({\rm quantum+inflow\  on\ I6})&
={3\over2}\Bigl(\tr F_{\sss SU_3}^2
-{1\over9}\tr R^2\Bigr)\cdot\Bigl(\tr F_{\sss E_6}^2+\tr F_{\sss SU_3}^2
-{1\over2}\tr R^2\Bigr),\cr
\noalign{\vskip 2\jot}
{\cal A}({\rm quantum+inflow\  on\ I6'})&
={3\over2}\Bigl(\tr F_{\sss SU_3}^2
-{1\over9}\tr R^2\Bigr)\cdot\Bigl(\tr F_{\sss SU_9}^2
-{1\over2}\tr R^2\Bigr).\cr
}}
As in our discussion in sect.~3, this anomaly can be cancelled
by an intersection-anomaly of the form \ansixs, provided that we
choose the parameters in $Y_4$ as $\eta={1\over9}$ and $\rho=1$.

\noindent
{\it $\IZ_4$ orbifold}

For this model the discussion is complicated by the fact that we have two
different types of fixed points. As we have explained before,
the local physics at the $\IZ_2$ fixed points is identical to
that of the $\IZ_2$ orbifold model of sects.~2 and 3. 
This in particular means that anomaly
cancellation on the $\IZ_2$ six-planes works in exactly the same way
as before. When computing the bulk and the ten-plane contribution to the
anomaly on the $\IZ_4$ six-planes, we must however first subtract
the contribution already accounted for on the $\IZ_2$ planes and then
distribute the remaining anomaly over the four $\IZ_4$  planes.

Before illustrating this for the charged ten-plane fields,
we will introduce some convenient notation.
We will denote the multiplet content of the charged ten-plane fields
which contribute to the anomaly by $Q_{10}$. This splits into
hyper-multiplets and vector-multiplets. Taking into account the opposite
chirality of the fermions in these multiplets, we write
$Q_{10}=H_{10}-V_{10}$.
The net number of states will be denoted by $n_{Q_{10}}$.
Let us consider the $SO_{6}\times SO_{10}$ side.
At the $\IZ_2$ fixed six-planes, the local physics is as in the
$\IZ_2$ model of sects. 2 and 3., {\it i.e.} the untwisted states consist
of a $SO_{16}$ vector-multiplet and one
$(\ul{128})_{SO_{16}}$ hyper-multiplet.
We now assume that the contribution to the anomaly of a $\IZ_2$ six-plane
is exactly that of such a six-plane in the $\IZ_2$ orbifold, {\it i.e.}
that of one sixteenth of a $SO_{16}$ vector-multiplet and 
of one sixteenth of a $\ul{128}$ hyper-multiplet.
Taking into account that there are six I6 of this type and four
I6 fixed under $\IZ_4$, we get for $Q_{10}$
\eqn\Qten{\eqalign{
Q_{10}&={1\over4}\left\lbrace\Bigl[(\ul{16},\ul{4})
-(\ul{45},\ul{1})-(\ul{1},\ul{15})
\Bigr]_{SO_{10}\times SO_6}
-{6\over16}\Bigl[(\ul{128})-(\ul{120})\Bigr]_{SO_{16}}
\right\rbrace\cr
&=-{5\over32}[(\ul{45},\ul{1})+(\ul{1},\ul{15})]
+{3\over32}(\ul{10},\ul{6})+{1\over16}(\ul{16},\ul{4})\,.}}
In the second step we have decomposed the $SO_{16}$ representation
under $SO_{16}\to SO_6\times SO_{10}$. Also,
$n_{Q_{10}}={1\over4}$.

It follows from the construction of the states in the untwisted
sector that all components of the decomposition of the
$(\ul{248})_{E_8}$ under $E_8\to SO_6\times SO_{10}$ have
definite $\IZ_4$ eigenvalue, namely
$e^{2\pi i\delta\cdot P}$ where $P$ is a $E_8$ root and
$\delta$ the shift vector in the $E_8$.
One finds
$\ul{248}=(\ul{15},\ul{1})_{+1}+(\ul{1},\ul{45})_{+1}
+(\ul{6},\ul{10})_{-1}+(\ul{4},\ul{16})_{+i}
+(\ul{\bar 4},\ul{\overline{16}})_{-i}\equiv \al(\ul{248})$. 
The subscripts are the $\IZ_4$ eigenvalues. 
Here $\al$ denotes the $\IZ_4$ generator whose action on a root of $E_8$
is specified by the shift vector.
Introducing the function
$T(x)={x\over8}+{x^2\over 32}$, whose argument can be either a complex
number or an operator, we can rewrite \Qten\ as
\eqn\QtenT{Q_{10}=-T(\al)(\ul{248})\,.}
The justification for introducing this notation is that
one can define a function $T(x)$ for all $\IZ_N$ orbifolds and
this function is universal for any given $N$, independent
of the choice of shift vector.
Specifically,
\eqn\talpha{
T({x})=\cases{{x\over16},& $N=2\,,$\cr
                  {x\over9}, & $N=3\,,$\cr
                  {x\over8}+{x^2\over32}, & $N=4\,,$\cr
                  {x\over6}+{x^2\over18}+{x^3\over 48}, & $N=6\,.$\cr}
}
We will need the values $T(1)=\lbrace{1\over16},{1\over9},{5\over32},
{35\over144}\rbrace$ and
$2 {\rm Re}\bigl(T(e^{2\pi i/n})\bigr)=\lbrace-{1\over8},-{1\over9},
-{1\over16},{5\over72}\rbrace$ for $n=\lbrace 2,3,4,6\rbrace$, respectively.
One checks that for every orbifold model we are considering,
$\sum_{\rm I6}T(1)=\sum_{{\rm I6}'}T(1)=1$ and
$-2\sum_{\rm I6}{\rm Re}(T(e^{2\pi i/n}))
=-2\sum_{{\rm I6}'}{\rm Re}(T(e^{2\pi i/n}))
={1\over2}\cdot\#$(moduli), where in the case of non-prime
orbifolds different types of fixed points
have to be summed over.

{}For the $SU_8\times SU_2$ side of the $\IZ_4$ orbifold one finds
in a similar way
\eqn\Qten{\eqalign{
Q_{10'}&={1\over4}\left\lbrace\Bigl[(\ul{28},\ul{2})
-(\ul{63},\ul{1})-(\ul{1},\ul{3})
\Bigr]_{SU_{8}\times SU_2}
-{6\over16}\Bigl[(\ul{56},\ul{2})-(\ul{133},\ul{1})
-(\ul{1},\ul{3})\Bigr]_{E_7\times SU_2}
\right\rbrace\cr
&=-{5\over32}[(\ul{63},\ul{1})+(\ul{1},\ul{3})]
+{3\over32}(\ul{70},\ul{1})+{1\over16}(\ul{28},\ul{2})\cr
&=-T(\al)(\ul{248})\,,}}
with $n_{Q_{10'}}=-{1\over4}$.

In addition to the ten-plane fields, on any six-plane
there is also the contribution
from the seven-plane fields with free boundary conditions on the
six-plane, again with a relative sign between hyper and vector multiplets.
In analogy to the notation introduced above,
we will denote them as $Q_7$. The subscript denotes the seven-dimensional
origin of these states.
Including a factor of ${1\over 2}$, which was explained in sect.~3 and
denoting the fields by their $SU_4^{\rm non-pert}$ representation,
we have for the $\IZ_4$ orbifold
$Q_7={1\over 2}\cdot\ul{15}$ and $Q_{7'}=-{1\over2}\cdot\ul{15}$.

As for the six-plane fields, there are none on the I6 planes. All
$\IZ_4$ twisted matter fields live on the I6$'$ planes.
{it I.e.} in the by now familiar notation,
$Q_6=\emptyset$ and $Q_{6'}=(\ul{1},\ul{4};\ul{8},\ul{1})
+{1\over2}(\ul{1},\ul{6};\ul{1},\ul{2})$.

If we define $Q=Q_{10}+Q_7+Q_6$ we can summarize the contribution
of all charged matter fields to the quantum anomaly as
\eqn\anQ{
{\cal A}(Q)=\Tr_{Q}F^4-{1\over4}\tr R^2\,\Tr_{Q} F^2
+n_{Q}\left({1\over240}\tr R^4+{1\over192}(\tr R^2)^2\right)\,.}

With the help of the function $T$ we can also express the
contribution to the anomaly of the bulk fields, namely
\eqn\anone{\eqalign{
{\cal A}(\hbox{SUGRA+tensor})&={1\over2}T(1)
\left[-{244\over240}\tr R^4+{44\over192}(\tr R^2)^2\right]\,,\cr
{\cal A}({\rm moduli})&=-2{\rm Re}(T(e^{2\pi i/n}))
\left[{1\over240}\tr R^4+{1\over 192}(\tr R^2)^2\right]\,.}}
Note {\it e.g.} that for $\IZ_4$ planes, $T(1)={1\over4}(1-{6\over16})$ and
$-2{\rm Re}(T_4(e^{2\pi i/n}))={1\over 2\cdot 4}(2-4\cdot{6\over16})$,
where for the latter we have taken into account that
a $\IZ_4$ orbifold has two moduli whereas a $\IZ_2$
orbifold has four moduli. \anone\ holds for both the I6 and the I6$'$
$\IZ_4$ planes.

The total quantum anomaly is thus
\eqn\totalQ{\eqalign{
{\cal A}({\rm quantum})&
=\Tr_Q F^4-{1\over4}\tr R^2 \Tr_Q F^2
+{1\over 240}\Bigl(n_Q-122\, T(1)-2\,{\rm Re}T(e^{2\pi i/N})\Bigr)\tr R^4\cr
&\qquad+{1\over 192}\Bigl(n_Q+22T(1)
-2{\rm Re}T(e^{2\pi i/N})\Bigr)(\tr R^2)^2\,.}}
This expression is valid for every $\IZ_N$ six-plane of any $\IZ_N$
orbifold, once $Q$ has been specified.

To compute the inflow contribution \anfives\ we need the magnetic
charge of the six-planes.
It can be determined from the
sum rule \gks.
We have to distribute the total charge $k$ over all
fixed planes on a given side of the $x^{11}$ interval.
For prime orbifolds they are all related by symmetry and carry the same
magnetic charge. For non-prime orbifolds some care is required.
{\it E.g.} on the
$SO_{10}\times SO_6$ side of the $\IZ_4$ model,
with total charge zero ($k_1=k_2=0$), there are
six $\IZ_2$ fixed planes with local gauge group $SO_{16}$, each
with charge ${1\over 4}$. The four $\IZ_4$ fixed planes must thus
carry a total charge of $-6\cdot{1\over4}$ or $-{3\over8}$ each.
{\it I.e.} $g_{\rm I6}=-{3\over8}$ and $g_{{\rm I6}'}=+{3\over8}$.

Note that there is a minimal magnetic charge any fixed plane must carry.
This is obtained if $k={-12}$, which corresponds to an unbroken $E_8$.
In this case there are no gauge instantons. It is straightforward to compute
the minimal magnetic charges of the various $\IZ_n$ planes.
They can be conveniently summarized in the formula
$g^{\rm min}_{\sss \IZ_n}={1-n^2\over 2n}$.
The allowed magnetic charges are then
$g_{\sss \IZ_n}=g^{\rm}+{m\over n}$ where the non-negative
integer $m$ counts the
number of $\IZ_n$ instantons sitting at the fixed point.

We have now provided all ingredients necessary to compute the
quantum and the inflow anomaly for the $\IZ_4$ model. A short
calculation gives
\eqn\qplusinzf{
\eqalign{
{\cal A}({\rm quantum+inflow\ on\ I6})&
={3\over2}\Bigl(\tr F_{\sss SU_4}^2
-{5\over32}\tr R^2\Bigr)\cdot\Bigl(\tr F_{\sss SO_{10}}^2+\tr F_{\sss SO_6}^2
-{1\over2}\tr R^2\Bigr),\cr
\noalign{\vskip 2\jot}
{\cal A}({\rm quantum+inflow\ on\ I6'})&
={3\over2}\Bigl(\tr F_{\sss SU_4}^2
-{5\over32}\tr R^2\Bigr)\cdot\Bigl(\tr F_{\sss SU_8}^2
+\tr F^2_{\sss SU_2}-{1\over2}\tr R^2\Bigr).\cr
}}
We once again find that this anomaly can be cancelled via \ansixs\ with
$\rho=1$ and $\eta=T(1)={5\over 32}$.

\noindent
{\it $\IZ_6$ orbifold}

We will be very brief here. We only have to check anomaly
cancellation on the $\IZ_6$ six-planes. The ten-plane gauge
groups are $SU_6\times SU_3\times SU_2$ on 10P and
$SU_9$ on 10P$'$. Using $T(x)$ as given in \talpha\
it is straightforward to show that
\eqn\Qtenzsix{\eqalign{
Q_{10}&=-T(\al)(\ul{248})\cr
&=-{35\over144}[(\ul{35},\ul{1},\ul{1})+(\ul{1},\ul{8},\ul{1})
+(\ul{1},\ul{1},\ul{3})]
-{5\over72}(\ul{6},\ul{3},\ul{2})+{13\over72}(\ul{15},\ul{\bar 3},\ul{1})
+{19\over 144}(\ul{20},\ul{1},\ul{2})\,,\cr
Q_{10'}&={13\over72}\cdot\ul{84}-{135\over144}\cdot\ul{80}}}
where the decomposition is with respect to the ten-plane
gauge groups. Also, the discussion in sect.~4.1 gave
\eqn\Qsixzsix{Q_6=\emptyset\,,\qquad
Q_{6'}=(\ul{6},\ul{1},\ul{1};\ul{\bar 9})
+{1\over2}(\ul{20},\ul{1},\ul{1};\ul{1})}
and
\eqn\Qsevenzsix{
Q_7={1\over2}\cdot\ul{35}\,\qquad Q_{7'}=-{1\over2}\cdot\ul{35}}
where the latter states are with respect to $SU_6^{\rm non-pert}$.
The magnetic charges are easily determined to
$g_{\rm I6}=-{5\over12}$ and $g_{{\rm I6}'}=+{5\over12}$.

For both the I6 and the I6$'$ planes, the quantum+inflow anomaly
can be cancelled via an intersection anomaly \ansixs\
with $\rho=1$ and $\eta=T(1)={35\over144}$.

\subsec{Common features}

We have demonstrated in all four examples that the local anomaly
on each intersection six-plane cancels.
This is to say that in the sum of quantum + inflow + intersection
anomaly the coefficients of $\tr R^4$, $(\tr R^2)^2$, $\tr R^2$ and
of the term without dependence on the Ricci-form vanish
separately. The four conditions for this to happen are:
\eqn\cancellation{\eqalign{
n_Q&=122 T(1)+T(e^{2\pi i/n})+T(e^{-2\pi i/n})+30 g\,,\cr
\eta&=T(1)\,,\cr
{1\over3}\Tr_Q F^2&=\rho\,\tr F_7^2+(g+2\eta)\tr F_{10}^2\,,\cr
{2\over3}\Tr_Q F^4&=\tr F_{10}^2\Bigl({g\over2}\tr F_{10}^2
+\rho\tr F_7^2\Bigr)\,.}}
In particular, local anomaly cancellation fixed the coefficients
$\eta$ and $\rho$ in \ysevens\ to $\eta=T(1)$ and $\rho=1$, respectively.
There is another way to see why these values are generic, which we will
now present. It also uses anomaly arguments and it involves a
direct comparison of the heterotic and the M-theory point of views.

In the heterotic theory, anomaly cancellation is of course
guaranteed by the well-established
consistency of the perturbative heterotic string. However, by realizing
that the massless fields which contribute to the heterotic anomaly on
a given fixed six-plane are precisely those which, in M-theory,
contribute on a I6, I6$'$ pair which is connected by a seven-plane,
we can, by comparison, determine the coefficients $\eta$ and $\rho$.

The boundary conditions of the seven-plane fields were chosen such that
in the limit $R_{11}\to 0$ there are
no additional massless states. This is reflected in
$Q_7+Q_7'=0$. Also, the magnetic charges satisfy
$g_{\rm I6}+g_{{\rm I6}'}=0$. We then have
$Q_{\rm net}\equiv Q+Q'=Q_{10}+Q_{10}'+Q_6+Q_6'$ and the first condition
in \cancellation\ gives
$n_{Q_{\rm net}}=244 T(1)+4\,{\rm Re}\,T(e^{2\pi i/n})$, which
is correct in the heterotic context without any reference to M-theory.

We can now compute the anomaly on a I6, I6$'$ pair from the heterotic and
from the M-theoretic point of view. The former gives ({\it cf}.\ Appendix~B)
\eqn\comb{\eqalign{
{2\over3}{\cal A}&={2\over3}\Tr_{Q_{{\rm net}}}F^4
-{1\over6}\tr R^2\Tr_{Q_{{\rm net}}}F^2
+T(1)(\tr R^2)^2\cr
&=(\tr R^2-\sum_\al\tr F_\al^2)\wedge
\Bigl(T(1)\tr R^2-\sum_\al\tilde v_\al\tr_\al F^2\Bigr)\,.}}
Computing the same anomaly in the M-theory picture produces instead
(use the last two equations in \cancellation)
\eqn\anpolpp{
{2\over3}{\cal A}=\Bigl(\tr R^2-\tr F_{10}^2-\tr F_{10'}^2\Bigr)\wedge
\Bigl(\eta\,\tr R^2+{g\over2}(\tr F_{10}^2-\tr F_{10'}^2)+\rho\,\tr F_7^2
\Bigr)\,.}
Comparison gives once more $\eta=T(1)$ and
\eqn\comparison{\eqalign{
v_{10}=v_{10'}=1\quad&\hbox{and}\quad
\tilde v_{10}={1\over 2}\sum_{\rm fixed\atop \rm planes}g
=-\,\tilde v_{10'}\,,\cr
v_{7}=0\quad&\hbox{and}\quad
\tilde v_7=\rho\,.}}
The seven-plane gauge groups $G_7$ thus have the characteristic of
non-perturbatively generated gauge groups. However, in the heterotic dual,
which is completely perturbative, they are not visible as additional
gauge group factors. This leaves two
options: $(i)$ $\rho\neq0$ and $\rho$ is the level of the gauge group $G_7$.
This is the situation we have encountered in all four examples considered
so far. The seven-plane gauge group mixes with the ten-plane gauge group,
as {\it e.g.} in \vtilde.
$(ii)$ $\rho=0$: this case will be discussed in sect.~5.

So far we have determined the parameters $\rho$ and $\eta$
using anomaly arguments. However, ultimately these parameters should
come from seven-dimensional physics. In fact, we will now
give an independent `derivation'
of the values for $\eta$ and $\rho$, which does not rely
on any anomaly arguments.

In M-theory on K3,
the parameters $\eta$ and $\rho$ enter through the electric coupling
of the seven-plane to the three-form potential $C$ of
11-dimensional supergravity via the term $C\wedge Y_4$, c.f.\ysevens\
and \FOL.
The eleven-dimensional origin of this term are the two
CS terms $C\wedge G\wedge G$ and $C\wedge(\tr R^4-{1\over4}(\tr R^2)^2)$.
However, it is easier to discuss these couplings from the dual
point of view, exploiting the duality between M-theory on K3 and
the heterotic theory on $T^3$ \Witten,\Sen.\foot{This seven-dimensional
heterotic theory is of course completely different from the Ho\v{r}ava--Witten
theory we have discussed so far.}

This heterotic -- M-theory duality in $d=7$ relates the field strength of
$C$, denoted by $G$, to the field strength $H$ of the Kalb-Ramond
field $B$ of the heterotic theory and vice versa:
$H\leftrightarrow *G$.
The moduli spaces of the heterotic compactification on $T^3$
and of M-theory on K3 are isomorphic.
At a generic point on the Narain lattice, the
gauge symmetry of the heterotic string compactified on $T^3$ is
$U(1)^{22}$ which is also the gauge group of the M-theory at a generic
point of the K3 moduli space.
On the heterotic side, the Bianchi identity reads
\eqn\bianchihet{
dH\propto\tr R^2+\sum_{I,J=1}^{22} d_{IJ} F_I F_J
}
where $d_{IJ}$ is a Lorentzian metric with signature $((+)^3,(-)^{19})$
which is also
the signature of the intersection matrix of the K3 homology 2-cycles.
Duality now implies $d(*G)\propto dH$ and thus an electric  coupling
$\propto C\wedge dH$.
At special points in the Narain moduli space the heterotic gauge symmetry
acquires non-Abelian components which contribute to $dH$, {\it i.e.}
\bianchihet\ gets modified to
\eqn\bianchihetmod{
dH\propto\tr R^2-\sum\tr F_i^2 ~+~\hbox{Abelian part}}
For the M-theory the gauge group enhancement
happens at the orbifold points of the K3 moduli
space. The additional states which comprise the non-Abelian gauge multiplets
are provided by M2 branes wrapping the vanishing cycles
(c.f. the discussion in sect.~1).
At the orbifold points the seven-dimensional gauge groups are
$SU_2^{16}\times U_1^6$ for the $\IZ_2$ orbifold,
$SU_3^9\times U_1^4$ for $\IZ_3$, $SU_2^6\times SU_4^4\times U_1^4$
for $\IZ_4$ and $SU_2^5\times SU_3^4\times SU_6\times U_1^4$
for $\IZ_6$.

{}From the eleven-dimensional point of view the Abelian part
lives in the bulk and is completely broken after compactification
to six-dimensions. It is thus of no further interest for us.
The non-Abelian part, on the other hand, contributes to the local
$C\wedge Y_4$ coupling on the seven-planes as
$-C\wedge \sum_i\tr F_{7i}^2$.
The sum is over all fixed seven-planes.
As far as
the $C\wedge \tr R^2$ piece of this coupling is concerned, it has
to be apportioned between all seven-planes. The apportioning happens
in exactly the same manner as with the bulk anomaly in previous sections,
namely as $T(1)\,\tr R^2$. The reason for this is the same as before.
The contribution on a $\IZ_N$ plane depends only on $N$, independent
of the orbifold model.

We thus find that the relevant coupling is proportional to
\eqn\relcoup{
C\wedge\Bigl(\tr R^2-\sum_i\tr F_{7i}^2\Bigl)
=C\wedge\sum_i\Bigl(T_i(1)\tr R^2-\tr F_{7i}^2\Bigl)}
Comparing this with $C\wedge\sum_i Y_{4i}$ leads to the
identification $\eta=T(1)$ and $\rho=1$.

We should point out 
that we have not been careful with the overall normalization of the
$C\wedge Y_4$ coupling. In fact, this coupling is somewhat controversial
since it is related to the normalization of the two
eleven-dimensional CS terms. Nevertheless the normalization must be universal
and it can be fixed by considering any orbifold model,
such as the $\IZ_2$ model of sects. 2 and 3.\foot{If one
changed the {\it relative} normalization of the two terms in
\anfives\ one would find that the second of the conditions in
\cancellation\ will be modified ($\eta$ and $T(1)$ will no longer
be proportional to each other),
in contradiction with
the independent arguments given here.}

\newsec{Open problems}

All the orbifold models we have discussed thus far followed a common
pattern from the M-theory point of view:
Each $\IZ_n$ fixed seven-plane carries a non-perturbative
$SU_n$ gauge theory which mixes with a perturbative gauge theory
living on one of the two ten-planes.
The resulting theory contains an $SU_n$ gauge group which appears
to be a subgroup of the $G\subset E_8$ (or $G'\subset E'_8$)
but isn't actually confined to one side of the $x^{11}$ interval.
Instead, it reaches to the other side along the fixed seven-planes
--- and that's how the twisted states living on the I$6'$ intersections
manage to have charges under both the $SU_n\subset G$ and the $G'$
gauge groups.

However, in many other orbifold models, the non-perturbative
gauge groups living on the $\IZ_n$ fixed seven-planes
turn out to be proper subgroups $G_7\subset SU_n$ rather that complete
$SU_n$'s.
Furthermore, such reduced non-perturbative groups --- or some of their
factors --- do not mix with the perturbative gauge group factors
but simply decouple from the massless states of the 
six-dimensional theory.\foot{%
    Technically, the locking boundary conditions~\lock\ at one side of
    the $x^{11}$ interval are replaced with the simple Dirichlet
    boundary  conditions $A^{\rm non-pert}_\mu(x^{11}=0)=0$.}
At present, we do not know any M-theoretical rules governing breaking
of the non-perturbative gauge groups or their mixing with the perturbative
gauge group factors.
All we have are the `experimental data' about the non-perturbative
gauge groups implied by the quantum numbers of the twisted states
in a score of orbifold models we have studied in some detail.

We see no point in boring the reader with technical details of too many models.
Instead, we shall simply present examples of two common problems we have
seen in several models.
In the following section 5.1
we discuss models with the unbroken $E'_8$ gauge group.
`Experimentally', all such models have $G_7=(U_1)^{(n-1)}$, the
Cartan subgroup of the $SU_n$, and none of the non-perturbative
$U_1$ factors mixes with any perturbative gauge groups.
Furthermore, the local anomaly cancellation requires the non-perturbative
$U_1$'s to have $\rho=0$, in blatant contradiction with the
seven-dimensional arguments of section~4.4.

In section 5.2 we present an example of a more complicated
model where the combined quantum, inflow and intersection anomalies 
{\sl do not cancel out} for any $G_7$ consistent with the
twisted states' quantum numbers.
This problem occurs in all models with mixed perturbative/non-perturbative
abelian gauge fields, although it may affect the non-abelian fields as well.
We speculate how seven-dimensional Chern--Simons
couplings may solve this problem, but a thorough analysis has to be postponed
to a later publication.

\subsec{Models with an unbroken $E_8$}

To exemplify the problems that arise in this class of models,
consider the best known heterotic orbifold, namely the $\IZ_2$
orbifold with the standard imbedding of the spin connection into
the gauge group, {\it i.e.}\ $\delta=(-{1\over2},+{1\over2},0,0,0,0,0,0;
0,0,0,0,0,0,0,0)$.
In this model, the first $E_8$ is broken down to $G=E_7\times SU_2$
while the second $E_8$ remains unbroken.
Note that we have already encountered such $\IZ_2$ fixed points
in the $\IZ_6$ orbifold in section 4, but let us take a closer look now. 
The hypermultiplet spectrum and other technical details of the `standard'
$\IZ_2$ orbifold are summarized in the following table:
%
%
$$
\vbox{\offinterlineskip\tabskip=0pt
\halign{\strut
\vrule width0.8pt \quad #
\tabskip=0pt plus 100pt
& \unskip\quad $#$ \hfill \quad
&\vrule#& \quad \hfill $#$ \hfill\quad
&\vrule#& \quad \hfill $#$ \hfill\quad
&\vrule width0.8pt#\cr
\noalign{\hrule}\noalign{\hrule}
& \multispan 5 \hfill Example 5:\quad$\IZ_2$ {orbifold with unbroken} 
$E_8$\hfill &\cr
\noalign{\hrule}\noalign{\hrule}
& \hbox{shift vector} && ({-1\over2},{+1\over2},0,0,0,0,0,0) &&
 (0,0,0,0,0,0,0,0)&\cr
\noalign{\hrule}
& \hbox{gauge group} && \hfill E_7\times SU_2 \hfill &&
\hfill E_8\hfill &\cr
\noalign{\hrule}
& \hbox{matter} \smash{\lower7pt
\hbox{$~~~H_0\quad\cases{\phantom{.}&\cr\phantom{.}&\cr
\phantom{.}&\cr}$}} \!\!\!\!\!\!\!\!\!\!\!\!\!\!\!\!
&& \multispan 3 \hfill\hbox{four moduli} \hfill &\cr
& &&
(\ul{56},\ul{2};\ul{1}) & \omit & \hbox{---} &\cr
& \hfill ~~~~~~~~~H_1 &&
\hbox{$16\times\lbrace{1\over2}(\ul{56},\ul{1};\ul{1})+
2(\ul{1},\ul{2};\ul{1})\rbrace$} &\omit& &\cr
\noalign{\hrule}
& b(G) && b(E_7)=42,\,\, b(SU_2)=42 &&  b(E_8)=-30 &\cr
& \tilde v^{~}_G && \tilde v^{~}_{E_7}=6,\, \tilde v^{~}_{SU_2}=6
&& \tilde v^{~}_{E_8}=-6 &\cr
& k=n-12 && +12 && -12 &\cr
& g^{~}_{\rm I6} && +{3\over 4} && -{3\over 4} &\cr
\noalign{\hrule}
&Q_{10} && {1\over 16}\lbrace(\ul{56},\ul{2})
-(\ul{133},\ul{1})-(\ul{1},\ul{3})\rbrace &&
-{1\over16}\cdot(\ul{248}) &\cr
&Q_6 && {1\over2}(\ul{56},\ul{1};\ul{1}^0)+2(\ul{1},\ul{2};\ul{1}^0)
&& \hbox{---} &\cr
\noalign{\hrule}
& G_7 && \multispan 3 \hfill\hbox{$U_1\subset SU_2$}\hfill &\cr
\noalign{\hrule}
&Q_7 && -{1\over2}\cdot\ul{1}^0&& {1\over 2}\cdot\ul{1}^0&\cr
\noalign{\hrule}
\noalign{\hrule}
}
\vskip.2cm
\noindent
{\ninepoint {\bf Table 5.1}: The $\IZ_2$ orbifold with the
$(E_7\times SU_2)\times E_8$ gauge group.}
}
$$

Note that unlike the `non-standard' $\IZ_2$ orbifold discussed in
section~2, the `standard' $\IZ_2$ orbifold has $\tilde v={k\over 2}$
for all gauge group factors including $SU_2$;
indeed, on the $E_7\times SU_2$ side, we have $k=+12$ and $\tilde v_{SU_2}=6$
rather than $\tilde v_{SU_2}=6+16=22$.
According to eq.~\vtilde, this indicates that in the standard $\IZ_2$ orbifold,
the $SU_2$ gauge factor is purely perturbative and does not mix with
any non-perturbative factors.
Consequently, no massless states in this model can be simultaneously charged
under the $SU_2$ and the $E_8$ gauge groups --- and indeed there are no
such states in the standard orbifold.

The non-perturbative gauge group $G_7$ on each of the 16 fixed seven-planes
must fit inside an $SU_2$ (bigger groups are not available at $A_1$
singularities such as $\IZ_2$ fixed points), but because it does not
mix with the perturbative $SU(2)$, we may have either $G_7=SU_2$ or $G_7=U_1$.
The choice of $G_7$ affects the $Q_7$ contribution to the quantum anomaly
at each end of the $x^{11}$: Given our general rule of opposite
boundary conditions on opposite ends (two avoid non-perturbative massless
states in 6d) and the requirement for free vector fields at either end
to form a closed subgroup of the $G_7$, it follows that
$n(Q^{(\prime)}_7)=\mp{3\over2}$
for $G_7=SU(2)$ but $n(Q^{(\prime)}_7)=\mp{1\over2}$ for $G_7=U_1$.
Since the $Q^{(\prime)}_6$ and $Q^{(\prime)}_{10}$
contributions to the anomaly on each side
are completely fixed by the massless spectrum of the orbifold, and since
$k=\pm12$ implies $g=\pm{3\over4}$, the need to cancel the $\tr R^4$ term
in the local anomaly at each end ({\it cf.}~first eq.~\cancellation) requires
$n(Q_7)=-{1\over2}$ on the $E_7\times SU_2$
side and $n(Q'_7)=+{1\over 2}$ on the $E_8$ side
--- and hence $G_7=U_1\subset SU_2$ rather than $G_7=SU(2)$.

Without going into any more details of eqs.~\cancellation, let us simply state
that for $G_7=U(1)$, the entire local anomaly polynomials $\cal A$ on both
I$6$ and I$6'$ planes cancel, provided (1) all the twisted states $Q_6$
are neutral with respect to the non-perturbative $U_1$ and (2) $\rho=0$
in the intersection anomaly term, all the arguments in section 4.4 for
$\rho=1$ notwithstanding; note however that $\eta={1\over16}=T(1)$ as
required by the second eq.~\cancellation.
At present, we have no explanation for the `experimental fact' of
 $\rho=0$ except  that its required to cancel the anomalies locally.
Likewise, we have no M-theoretical explanation for the $G_7=U_1$,
only the brute fact that this too is required to cancel the anomalies locally.
The theory would have to wait for a later publication.

In lieu of theory, we offer a summary of `experimental' data to show a common
pattern of {\it invisible} non-perturbative abelian gauge groups,
which do not mix with any perturbative gauge group factors and don't couple
to any twisted massless states ({\it i.~e.}, the hypermultiplets living
on I$6$ and I$6'$ are neutral with respect to the invisible groups).
In particular, in {\sl all} $\IZ_n$ orbifolds ($n=2,3,4,6$) with an unbroken
$E_8$ group\foot{There are ten distinct heterotic orbifolds of this type,
    including the standard $\IZ_2$ orbifold, two different $\IZ_3$ orbifolds,
    two $\IZ_4$ orbifolds and five $\IZ_6$ orbifolds.}
we can cancel all anomalies locally, provided:
(1) each $G_7(\IZ_n\,{\rm plane})=(U_1)^{(n-1)}$, the Cartan subgroup of the $SU_n$;
(2) none of the non-perturbative $U_1$'s mixes with any of the perturbative
groups; (3) all twisted states are neutral under all the non-perturbative
$U_1$'s; and (4) all the non-perturbative $U_1$'s have $\rho=0$.
In addition, many $\IZ_4$ and $\IZ_6$ orbifolds have a locally-unbroken $E_8$
on the $\IZ_2$ or $\IZ_3$ fixed planes --- and all such fixed planes
carry invisible $U_1$ or $(U_1)^2$ gauge groups, exactly like similar fixed
planes in the corresponding $\IZ_2$ or $\IZ_3$ models.
For example, the $\IZ_6$ orbifold of section~4 has such an invisible
$G_7=U_1$ on each of  its $\IZ_2$ fixed planes.

It is easy to see that any invisible $U_1$ gauge field must have $\rho=0$
for the sake of local anomaly cancellation.
Indeed, an invisible seven-plane $U_1$ has no quantum anomalies
--- since all the chiral fields are neutral --- and no inflow anomalies
--- which involve only the ten-plane groups or the {\sl visible}
seven-plane groups that mix with them.
Consequently, any invisible $U_1$ factor should also decouple from the
intersection anomaly, which means it must have $\rho=0$.
Unfortunately, this anomaly-counting argument does not tell us how the
invisible seven-plane gauge groups differ from the visible seven-plane
groups {\sl from the seven-dimensional point of view}, so the M-theoretical
origins of the $\rho_{\rm invisible}=0$ remain obscure.

\subsec{Models with perturbative $U_1$ factors}

The invisible abelian groups of the previous section are puzzling
from the M-theoretical point of view, but as far as the local
anomaly cancellation is concerned, the {\it visible} abelian groups
are much more troublesome.
Generally, the combined quantum, inflow and intersection anomalies
in such models cancel each other globally but not locally in $x^{11}$,
\eqn\schiso{
{\cal A}({\rm I6})\ +\ {\cal A}({\rm I6}')\ =\ 0\quad
{\rm but}\quad {\cal A}({\rm I6})\ \neq\ 0\ \neq\ {\cal A}({\rm I6}').
}
In other words, the anomalies cancel locally in ten dimensions
(which always works in any perturbative heterotic orbifold) but not
the eleventh dimension.

As an example of such anomaly trouble, let us consider a $\IZ_4$ orbifold
summarized in the following table:

$$
\hskip 0pt minus 0.5in
\vbox{\offinterlineskip \ninepoint
\ialign{%
\strut \vrule width 0.8pt #\quad &
    #\unskip\hfil & \vrule #\quad &
    \hfil ${#}$\hfil\quad & \vrule #\quad &
    \hfil ${#}$\hfil\quad & \vrule width 0.8pt #\hfil\cr
\noalign{\hrule height 0.8pt}
& \multispan 5 \hfil
    Example 6:\quad $\IZ_4$ orbifold \hfil &\cr
\noalign{\hrule height 0.8pt}
& shift vector &
& (-{3\over4},{1\over4},{1\over4},{1\over4},0,0,0,0) &
& (-{1\over2},{1\over4},{1\over4},0,0,0,0,0) &\cr
\noalign{\hrule}
& gauge group &
& SO_{10}\times SU_4 &
& E_6\times SU_2\times U_1 &\cr
\noalign{\hrule}
& matter  $H_0\,\left\lbrace\vrule height 1em depth 1em width 0em\right.$&
& \vcenter{\kern 1em \hbox{$(\ul{16},\ul{4};\ul{1},\ul{1},0)$}}
& \omit $\vcenter{\hbox to 0pt{\hss two moduli\hss }\kern 1em}$\hss\quad
& \vcenter{\kern 1em \hbox{$(\ul{1},\ul{1};\ul{27},\ul{2},{-1\over\sqrt{12}})\,
                         +\,(\ul{1},\ul{1};\ul{1},\ul{2},{+3\over\sqrt{12}})$}} &\cr
&\phantom{matter}  $H_1$ &
& \multispan 3 \hfil
    $4\times \left\{(\ul{1},\ul{4};\ul{1},\ul{2},{+3\over2\sqrt{12}})\,
    +\,2(\ul{1},\ul{4};\ul{1},\ul{1},{-3\over2\sqrt{12}})\,
    +\,(\ul{\overline{16}},\ul{1};\ul{1},\ul{1},{+3\over2\sqrt{12}})\right\}
    $\hfil\quad &\cr
&\phantom{matter}  $H_2$ &
& \multispan 3 \hfil
    ${10\over 2}\times(\ul{10},\ul{1};\ul{1},\ul{2},0)\,
    +\,{6\over2}\times(\ul{1},\ul{6};\ul{1},\ul{2},0)
    $\hfil\quad &\cr
\noalign{\hrule}
& $b(G)$ &
& b(SO_{10})=b(SU_4)=+18 &
& \! b(E_6)=-6,\ b(SU_2)=+54,\ b(U_1)=+30\! &\cr
& $k=n-12$ && +4 && -4 &\cr
& $\tilde{v}_G$ &
& \tilde{v}_{SO_{10}}=\tilde{v}_{SU_4}=+2&
& \tilde{v}_{E_6}=-2,\quad \tilde{v}_{SU_2}=+8,\quad
    \tilde{v}_{U_1}=+4 &\cr
& $\delta\tilde{v}=\tilde{v}-{k\over 2}$ &
& \delta\tilde{v}_{SO_{10}}=\delta\tilde{v}_{SU_4}=0&
& \! \delta\tilde{v}_{E_6}=0,\quad \delta\tilde{v}_{SU_2}=6+4,\quad
    \delta\tilde{v}_{U_1}=4\times{3\over2}\! &\cr
\noalign{\hrule height 0.6pt}
& \multispan 5 \hfil $\IZ_2$ fixed points: See sections 2 and 3\hfil\quad &\cr
\noalign{\hrule height 0.6pt}
& \multispan 5 \hfil $\IZ_4$ fixed points\hfil\quad &\cr
\noalign{\hrule height 0.6pt}
& $g_{I6}$ && +{5\over8} && -{5\over8} &\cr
\noalign{\hrule}
& $Q_{10}$ &
& -{5\over32}\left\{(\ul{45},\ul{1})+(\ul{1},\ul{15})\right\} &
& -{5\over32}\left\{(\ul{78},\ul{1},0)+(\ul{1},\ul{3},0)+(\ul{1},\ul{1},0)\right\} &\cr
&&& {}+{1\over16}(\ul{16},\ul{4}) &
& {}+{1\over16}\left\{(\ul{27},\ul{2},{-1\over\sqrt{12}})
       +(\ul{1},\ul{2},{+3\over\sqrt{12}})\right\} &\cr
&&& {}+{3\over32}(\ul{10},\ul{6}) &
& {}+{3\over32}(\ul{27},\ul{1},{2\over\sqrt{12}}) &\cr
& $Q_6$ &
& (\ul{1},\ul{4};\ul{2},{+3\over2\sqrt{12}})\,
    +\,2(\ul{1},\ul{4};\ul{1},{-3\over2\sqrt{12}}) &
& \hbox{---} &\cr
&&& {}+\,(\ul{\overline{16}},\ul{1};\ul{1},{+3\over2\sqrt{12}})\,
    +\,{1\over2}(\ul{10},\ul{1};\ul{2},0) &&&\cr
\noalign{\hrule}
& $G_7$ &
& \multispan 3 \hfil
    $G_7\supset SU_2\times U_1$;\qquad $G_7\subset SU_4$\hfil\quad &\cr
\noalign{\hrule}
& $Q_7$ && \hbox{????} && \hbox{????} &\cr
\noalign{\hrule height 0.8pt}
}
\vskip 0.2cm
\lineskip=2pt plus 1 pt
\noindent {\bf Table 5.2}: A $\IZ_4$ orbifold with the
$(SO_{10}\times SU_4)\times (E_6\times SU_2\times U_1)$
gauge group and anomaly troubles.\par
}\hskip 0pt minus 0.5in
$$
Note that all data in this table are completely determined by the
perturbative heterotic string theory --- except of course for
the inherently non-perturbative $G_7$ and $Q_7$.
However, the model must have $G_7\supset SU_2\times U(1)$ since the
gauge couplings of the perturbative $SU_2$ and $U_1$ group factors
indicates their mixing with the non-perturbative gauge fields living
on the $\IZ_4$ fixed seven-planes.
Actually, the $SU_2$ factor gets non-perturbative contributions from
both $\IZ_2$ and $\IZ_4$ fixed planes, hence $\tilde{v}_{SU_2}-{k\over2}=6+4$.
In the $U_1$ case however, only the $\IZ_4$ fixed seven-planes are involved,
but there appear to be a non-trivial mixing angle, thus
$\tilde{v}_{U_1}=4\times{3\over2}$ rather than simply~4.

The question at this point is whether any consistent choice of $G_7$
and $Q^{(\prime)}_7$ would lead to a local cancellation of all the anomalies
on the I$6$ and I$6'$ planes.
The answer turns out to be negative in an interesting way:
The requirement of local anomaly cancellation does have a unique solution
\eqn\Tantalus{
Q_7\ =\ -Q'_7\ =\ -{1\over2}(\ul{3},0)\ -\ (\ul{2},{3\over\sqrt{12}}).
}
In terms of the seven-plane fields and their boundary conditions,
this means $G_7=SU_3\times U'_1\subset SU_4$
where $SU_3\supset SU_2\times U_1$ while the $U'_1$ factor is invisible.
In $SU_3\times U'_1$ terms, $Q_7={1\over2}\left\{\ul{1}^0-\ul{8}^0\right\}$,
which means free (Neumann) boundary conditions for all eight of $SU_3$ vector
fields at I$6$ on the $SO_{10}\times SU_4$ end of $x^{11}$.
Consequently, the fields living on the I$6$ intersection plane should
form complete multiplets of the $(SO_{10}\times SU_4)^{\rm pert}\times
SU_3^{\rm n.p.}$ gauge group visible at I$6$.
Unfortunately, they don't --- {\it cf.}\ the $Q_6$ row of the table~5.2 ---
which means the solution~\Tantalus\ to the local anomaly problems is
inconsistent with group theory.
Group-theoretically, the only possibilities consistent with all the
quantum numbers are $G_7=SU_2\times U_1$ and $G_7=SU_2\times U_1\times U'_1$.
Both choices lead to non-zero net gauge, gravitational 
and mixed local anomalies
on both the I$6$ and the I$6'$ intersection planes.

This type of local anomaly mis-cancellation is common in orbifolds with
perturbative abelian groups.
Generally, the quantum numbers of the twisted states in such models
require some fixed seven-planes to carry {\sl abelian} non-perturbative
gauge fields that lock onto the perturbative $U_1$ fields on one side
of the eleventh dimension, say at I$6$, but have free boundary conditions
on the other side at I$6'$.
Locally on the I$6'$ plane, there are no $F^4_{U_1}$ inflow or intersection
anomalies (since the $U_1$ does not lock onto the ten-plane gauge fields
on {\sl that} side) but there is a non-zero quantum $F^4_{U_1}$ anomaly
because there are $U_1$ charged hyper-multiplets (twisted states) but no
charged vector-multiplets (if there were charged vector fields, the
group would not be abelian).
Altogether, {\sl the abelian gauge anomaly $F^4_{U_1}$ does not cancel locally}
on the I$6'$ plane.

`Experimentally', in all the models we have studied, a mis-cancelled
abelian gauge anomaly is accompanied by mis-cancelled anomalies
involving other non-perturbative gauge fields living on the same plane
as well as gravity.
On the other hand, all anomalies involving the un-mixed purely perturbative
gauge fields do cancel out.
Algebraically,
\eqn\ExCS{
{\cal A}({\rm I6})\ =\ -{\cal A}({\rm I6}')\
=\ P(F_7,R)
}
for some quartic polynomial $P$ in seven-dimensional gauge
and gravitational curvature two-forms $F_7$ and $R$.

The algebraic structure~\ExCS\ of the net local anomaly suggests
that it may ultimately cancel against a Chern--Simons action term
in the effective theory on the fixed seven-plane.
Indeed, consider
\eqn\CSaction{
{\rm Action}\ \supset \int\limits_{7P}\! I_7
}
where $I_7$ is a Chern--Simons 7--form which varies under gauge
transformations by a total derivative, $\delta I_7=d I_6^1$.
Consequently,
\eqn\CSvariance{
\delta{\rm Action}\ = \int\limits_{\partial(7P)}\! I^1_6\
= \int\limits_{\rm I6'} I^1_6\ - \int\limits_{\rm I6} I^1_6\,,
}
which looks {\it exactly} like local anomalies on the I$6$
and I$6'$ boundary six-planes with exactly opposite anomaly
polynomials
\eqn\CSpoly{
I_8({\rm I6}')\ =\ - I_8({\rm I6})\ = dI_7\,.
}
Therefore, if $dI_7=P(F_7,R)$ ({\it cf.}~eq.~\ExCS)
or rather $dI_7=P/(192\pi^3 i)$, then all anomalies would cancel
locally.

We are currently investigating whether such Chern--Simons terms
do actually cancel the residual local anomalies of orbifold models
with abelian gauge factors.
There are (at least) two issues that must be addressed.
First, there is a question of normalization:
gauge, gravitational and mixed Chern--Simons terms all have quantized
coefficients, which may turn out to be consistent or inconsistent
with the local anomaly cancellation in various models.
The second open problem is the physical origin of
the 7d Chern--Simons terms.
All anomalies cancel locally in models considered in sections 2--4
without any help from the CS terms, so why should the CS terms appear
in other models?
At the moment the CS terms still seem to pose as many new questions
as they are able to answer.
We hope to be able to report some progress on this issue in the near future.

\newsec{Summary and Discussion}

In this paper we have investigated the correspondence between
perturbative heterotic $E_8\times E_8$ orbifolds on $T^4/\IZ_N,\,(N=2,3,4,6)$
and their M-theory duals.
Our aim was to obtain information about the structure
of M-theory orbifolds, in particular the spatial locations of all the charged
fields and the gauge groups that act upon them.
Our main tools were constraints arising from local consistency of quantum
numbers, from the gauge coupling
considerations and from the requirements of local anomaly cancellation.
We have put forward
a scenario which allowed us to resolve an apparent paradox associated with
the existence of twisted massless states simultaneously charged under gauge groups
which appear to live on two different boundary ten-planes.
The resolution relies on the mixing of a factor (or factors) of
the gauge group $G$ living on one boundary ten-plane with the
$G_7$ gauge groups living on the fixed seven-planes
$R^{5,1}\otimes\{{\rm f.p.}\}\otimes S^1/\IZ_2$ of the orbifold.
The twisted states which live on the I$6'$ planes where the fixed seven-planes
reach the other ten-plane boundary thus have perfectly local $G_7\times G'$
charges.
Thanks to mixing, however, the $G_7$ charges of those states
masquerade as the perturbative
$G$ charges from the six-dimensional point of view, hence the appearance
of the simultaneous $G$ and $G'$ charges in the heterotic perturbation theory.

Considerations of the gauge couplings confirm that there is a mixing 
between the non-perturbative gauge groups on the seven-planes and the 
perturbative gauge group on the boundary planes.
Another confirmation comes from the overall 6d anomaly polynomial.
The most powerful test however is the local anomaly cancellation.
Here the correct allocation of all states ---
in particular the hyper and vector component of the seven-plane gauge 
multiplets --- to the six-planes where the anomaly is required to 
cancel is absolutely crucial.
 A consistent assignment of states was only possible 
after a judicious choice of boundary conditions on the seven-plane 
fields which projects out the vector components on one side and 
the hyper-component on the other. This also guaranteed that there 
are no additional massless states in the $R_{11}\to 0$ heterotic
limit. 

There are three contributions to the net local anomaly: quantum anomalies, 
inflow anomalies and intersection anomalies.
Given a correct identification of all the locally relevant fields,
the quantum one-loop anomalies are fairly straightforward,
but the inflow and intersection anomalies involve un-settled issues
of the overall normalization.
In principle, the overall normalization of the inflow anomalies
follows from that of the Chern--Simons terms in eleven-dimensional SUGRA,
but the derivation is rather subtle and the result is controversial.
Instead, we simply calibrated the normalization by requiring local
anomaly cancellation in the $\IZ_2$ model ({\it cf.}\ section~3), then used
the same normalization in all the other models of section~4;
eventually, this normalization ought to be confirmed by a direct derivation.

Likewise, the $\eta$ and $\rho$ parameters of the intersection anomaly
({\it cf.}\ eqs.\ \ansixs and \ysevens) should follow from the
seven-dimensional Chern--Simons terms (which in turn follow from the M-theory)
and indeed the analysis of section 4.4 gives specific values $\eta=T(1)$
and $\rho=1$.
Alternatively, we can treat them as free parameters, fixed by the anomaly
cancellation requirements of each model.
According to eqs.~\cancellation, $\eta=T(1)$ is indeed universally valid,
but the $\rho$ parameter is more tricky:
$\rho=1$ works for model with $G_7=SU_n$ ({\it cf.}\ sections~3 and~4)
but models with the invisible abelian $G_7=U_1^{(n-1)}$
({\it cf.}\ section~5.1) require $\rho=0$ instead.
Unfortunately, we don't know yet the M-theoretic origin of the seven-plane
$SU_n$ breakdown to its Cartan subgroup in these models, and we don't know
how this Cartan subgroups ends up with $\rho=0$ either.

A worse anomaly trouble plagues models with abelian factors in their
perturbative gauge groups, {\it cf}.\ section~5.2.
In those models, the local anomalies simply do not cancel for any
consistent choices of the seven-plane gauge fields and their boundary
conditions at the two ends of the $x^{11}$.
This presumably means one of the two things:
Either we don't know how to correctly interpret such models
in M-theory terms or else there must be additional sources of local
anomalies.
For example, a Chern--Simons term in the seven-plane effective action
has an effect of transferring local 6d anomaly from one boundary of
the seven plane to the other boundary, thus possibly helping to cancel
the local anomalies on both boundaries.
Much work is needed however before we know whether this mechanism
really works.

Our main conclusion is that the  heterotic orbifolds do have consistent
M-theory duals with identical massless spectra and gauge couplings
and locally cancelled anomalies.
Unfortunately, while the duality and the local anomalies tell us
{\sl what} should happen on every six-, seven- or ten-plane of the
dual model, we do not understand {\sl how} it happens in M-theory.
The big mystery is the dynamic origin of the boundary
conditions for the seven-plane fields.
In particular, we know that each seven-plane field always has exactly
opposite boundary conditions on the two boundaries of the $7P$,
but the M-theoretic reasons for this twist remain completely obscure.
Another mystery why the $\IZ_n$-fixed seven-planes carry full $SU_n$
gauge groups in some models while in other models the $SU_n$ is broken
to a subgroup;
in fact, we are not even sure of the mechanism of such seven-plane symmetry
breaking.
We are however quite certain that the eventual resolution of these
issues will shed much new light on the basic structure of the M-theory.

We conclude this article with a few comments on M-theory duals of the
four-dimensional heterotic orbifolds.
Generally, such models live on
$\IR^{3,1}\otimes(T^6/\Gamma)\otimes (S^1/\IZ_2)$ comprising
the 11d bulk, fixed seven-planes, fixed five-planes as well as their
respective ten-plane, six-plane and four-plane boundaries.
The bulk, the seven-planes and their boundaries should behave just as they
do in the 6d orbifolds (modulo orbifolding of their extra compact coordinates),
it's the five-planes that radically complicate the physics.
The problem with the fixed five-planes is that they are poorly understood
from the M-theory point of view.
We know that the singular five-planes in M-theory carry superconformal
theories whose excitations include both massless particles and
tensionless strings.
Generally, the superconformal theories in five dimensions are associated
with infinite-coupling gauge theories, but the M-theory tells us nothing about
the gauge group, only the global symmetries (if any) and some
gauge-invariant operators --- and these data are quite insufficient
for the model building purposes.
Also, the constraints of local anomaly cancellation on the four-plane
boundaries of the five-planes are rather weak, simply because the
anomalies in four dimensions are much simpler than in six.
Consequently, although we have tentatively identified the M-theory
duals of a few 4d heterotic orbifolds, we do not have enough constraints
to be confident in our identification at the time of this writing.
This work is in progress and we hope to present some interesting results
in the near future.

\bigskip\noindent
{\bf Acknowledgements:}
An important part of the project was accomplished
while V.~K., J.~S. and S.~Y. visited  Universit\"at M\"unchen.
They  would like to thank the Sektion Physik
for its hospitality.  S.Y would also like to thank the theory group of CERN.
S.T. would like to thank R. Minasian and C. Scrucca for 
helpful discussions.
       
\vfill\eject
\noindent
\appendix{A}{Anomaly polynomials for ${\cal N}=1$ multiplets in d=6}

This appendix summarizes the six-dimensional gravitational,
gauge and mixed anomaly contributions of
all relevant six-dimensional ${\cal N}=1$ multiplets,
see {\it e.g.}~\GSW.
The representation content of each such multiplet is spelled out
in terms of the little group $SO_4\simeq SU_2\times SU_2$ for massless
particles in 6d.
The correctly normalized anomaly polynomials are
\eqn\cornorm{
I_8\ =\ {-i\over 4!(2\pi)^3}\,{\cal A},
}
but since we use $\cal A$ instead of $I_8$ throughout this paper,
the anomalies below are written in terms of $\cal A$ as well.

\noindent
$\bullet$~gravity multiplet $[(3,3)+2(2,3)+(1,3)]$:
\eqn\gravity{
{\cal A}_{\rm grav}=-{273\over240}\tr R^4
+{51\over192}(\tr R^2)^2\,.}
$\tr$ is the trace in the vector representation of $SO(6)$.

\noindent
$\bullet$~vector-multiplets $n_V[(2,2)+2(1,2)]$:
\eqn\vector{
{\cal A}_{\rm V}=-{n_{\rm V}\over240}\tr R^4
-{n_{\rm V}\over192}(\tr R^2)^2 +{1\over 4}\tr R^2 \Tr_{\rm V} F^2
-\Tr_{\rm V} F^4\,;}
since the vector multiplets comprise the adjoint representation of the
gauge group~$G$, $\Tr_V$ is the trace in the adjoint representation and
$n_V=dim(G)$. 

\noindent
$\bullet$~hyper-multiplets $n_H[2(2,1)+4(1,1)]$:
\eqn\hyper{
{\cal A}_{\rm H}={n_{\rm H}\over240}\tr R^4+
{n_{\rm H}\over192}(\tr R^2)^2 -{1\over 4}\tr R^2 \Tr_{\rm H} F^2
+\Tr_{\rm H} F^4\,,}
where $\Tr_{H}$ is the trace in the representation of $G$ comprised
of all the hypermultiplets (whatever such representation might be in any
particular model)
and $n_H$ is the net number of hypermultiplets in the model.

\noindent
$\bullet$~self-dual-tensor multiplet $[(3,1)+2(2,1)+(1,1)]$:
\eqn\tensor{
{\cal A}_{\rm tensor}={29\over240}\tr R^4
-{7\over192}(\tr R^2)^2\,.}

\bigskip

\noindent
\appendix{B}{Further anomaly considerations}

In section~2.2, we mentioned that in perturbative K3 compactifications
of the heterotic string, the gauge couplings are related to the 6d
anomaly polynomial.
In this appendix, we explain this relation.

The anomaly we are concerned with here is the overall anomaly of the
6d effective theory rather than local anomalies on some singular six-planes.
In perturbative K3 compactifications, there is just one tensor field $B_{\mu\nu}$,
hence the Green--Schwarz mechanism of anomaly cancellation works {\sl only}
if the net anomaly happens to factorize in the form~\DMW:
\eqn\factorize{
{2\over3}{\cal A}=(\tr R^2-\sum_\alpha v_\al\tr F_\al^2)\wedge
(\tr R^2-\sum_\al\tilde v_\al\tr F_\al^2)\equiv I_4\wedge\tilde I_4\,.}
Furthermore, the ${\cal N}=1$, $d=6$ supersymmetry relates the couplings of
the $B_{\mu\nu}$ tensor fields to those of the heterotic dilaton $\phi$,
hence the factorization coefficient $v_\al$ and $\tilde v_\al$ in this
formula also show up in the exact equations~\gaugekin\ for the gauge couplings.

The Green--Schwarz mechanism does not cancel the irreducible $\tr R^4$
term in the anomaly polynomial, hence anomaly-free theories with just one
self-dual-tensor multiplet must have $n_H-n_V=244$, {\it cf.}\ individual
multiplets' anomalies listed in Appendix~A.
Consequently, the net anomaly can be summarized as
\eqn\anpol{
{2\over3}{\cal A}=(\tr R^2)^2-{1\over6}\tr R^2\,\Tr_{H-V}F^2
+{2\over3}\Tr_{H-V} F^4\,,}
were the notation $\Tr_{H-V}$ denotes the trace taken over
all hyper-multiplets minus the trace over all
vector-multiplets.
The relative minus-sign follows from opposite chiralities of fermions in
${\cal N}=1$, $d=6$ hyper- and vector-multiplets.
Comparing the mixed gauge/gravitational anomaly terms in eqs.~\anpol\
and \factorize, we see that Green--Schwarz anomaly cancellation requires
\eqn\BVorigin{
\Tr_{H-V}F^2\ = \sum_\al 6(v_\al+\tilde v_\al)\,\tr F^2_\al\,.
}
(See Appendix~C for exact normalizations of various traces.)
Curiously, $\Tr_{H-V}F^2$ also appears in beta-function coefficients of
${\cal N}=2$, $d=4$ gauge theories (including but not limited to toroidal
compactifications of the ${\cal N}=1$, $d=6$ theories); specifically,
$\Tr_{H-V}F^2=\sum_\al b_\al\tr F^2_\al$.
Comparing this formula to eq.~\BVorigin, we immediately arrive at eq.~\bv.

{}From the M-theory point of view, there is another interesting way
to write the anomaly polynomial as a sum of two factorized terms, each
associated with the boundary at the corresponding end of $x^{11}$.
For smooth K3 compactifications (perturbative or otherwise) with
instanton numbers $n_1+n_2=24$ (hence exactly one self-dual-tensor
multiplet), we have~\SW
\eqn\twoterm{\eqalign{
{2\over3}{\cal A}&=\Bigl({1\over2}\tr R^2
-\sum_\al v_{\al 1}\tr F_{\al 1}^2\Bigr)
\wedge\Bigl({1\over4}(n_1-8)\tr R^2
-\sum_{\al}\tilde v_{\al 1}\tr F_{\al 1}^2\Bigr)\cr
&+\Bigl({1\over2}\tr R^2-\sum_\al v_{\al 2}\tr F_{\al 2}^2\Bigr)
\wedge\Bigl({1\over4}(n_2-8)\tr R^2
-\sum_{\al}\tilde v_{\al 2}\tr F_{\al 2}^2\Bigr)\,.}}
There are similar expressions for singular K3's such as orbifolds.
For example, the $\IZ_2$ model of sect.~2 has
\eqn\twoterms{\eqalign{
{2\over3}{\cal A}&=\Bigl({1\over 2}\tr R^2-\tr F_{E_7}^2-\tr F_{SU_2}^2\Bigr)
\wedge(2\tr F_{E_7}^2-14\tr F_{SU_2}^2\Bigr)\cr
&+\Bigl({1\over 2}\tr R^2-\tr F_{SO_{16}}^2\Bigr)
\wedge(2\tr R^2-2\tr F_{SO_{16}}^2-16\tr F_{SU_2}^2\Bigr)\,.}}
We observe that one of the gauge factors, namely
$SU_2$, now appears on both sides but only
non-perturbatively ($v=0,\,\tilde v=16$) on the $SO_{16}$ side.
This lends support to the M-theory description of this model
that we have put forward in sect.~2.
We have a non-perturbatively generated gauge group $(SU_2)^{16}$.
Each of these $SU_2$'s has $\tilde v=1$ and
they mix with the perturbative $SU_2$ such that only the diagonal
$SU_2$ contributes in the heterotic description of the model, {\it i.e.} only
one $SU_2$ is visible. This $SU_2$ is however visible on both
sides of the $x^{11}$ interval.

\bigskip

\noindent
\appendix{C}{Some results from group theory}

We collect some group theoretical results which are needed to
verify anomaly cancellation. The notation is such that
$\Tr_R$ is always the trace in the representation $R$ whereas
$\tr F^2$ means $\sum_a F^a F^a$. The normalizations are such that
the long roots are normalized to length 1.

$\underline{SU_2}$
\eqn\sutwo{
\eqalign{
\Tr_{\ul 2}F^2&\equiv{1\over2}\tr F^2\cr
\Tr_{\ul 2}F^4&={1\over8}(\tr F^2)^2\cr
\noalign{\vskip.2cm}
\Tr_{\ul 3}F^2&=2\tr F^2\cr
\Tr_{\ul  3}F^4&=2(\tr F^2)^2}
}

$\underline{SU_3}$
\eqn\suthree{
\eqalign{
\Tr_{\ul 3}F^2&={1\over2}\tr F^2\cr
\Tr_{\ul 3}F^4&={1\over8}(\tr F^2)^2\cr
\noalign{\vskip.2cm}
\Tr_{\ul 8}F^2&=3\tr F^2\cr
\Tr_{\ul 8}F^4&={9\over4}(\tr F^2)^2}
}

$\ul{SU_N,\,N\geq4}$
\eqn\sun{
\eqalign{
{}\quad &\Tr_{\ul N} F^2={1\over2}\tr F^2\cr
        &\Tr_{\ul N} F^4= {3\over16}(\tr F^2)^2-{1\over4}\tr F^4\cr
\noalign{\vskip.2cm}
{}\quad &\Tr_{\rm ad} F^2 = N\tr F^2\cr
        &\Tr_{\rm ad} F^4 ={3\over8}(N+4)(\tr F^2)^2-{N\over2}\tr F^4\cr
\noalign{\vskip.2cm}
\quad & \Tr^{~}_\Dbox F^2 = {N-2\over 2}\tr F^2\cr
      & \Tr^{~}_\Dbox F^4 = {3\over 16}(N-4)(\tr F^2)^2
                 +(2-{N\over4})\tr F^4\cr
\noalign{\vskip.2cm}
\quad & \Tr^{~}_\Tbox F^2={1\over4}(N-2)(N-3)\tr F^2\cr
      & \Tr^{~}_\Tbox F^4={3\over32}(N^2-9N+22)(\tr F^2)^2
           -{1\over8}(N^2-17 N+54)\tr F^4}
}

$\underline{SO_N,\, N\geq5}$
\eqn\son{
\eqalign{
{}\quad & \Tr_{\ul N} F^2\equiv\tr F^2\cr
        & \Tr_{\ul N} F^4\equiv \tr F^4\cr
\noalign{\vskip.2cm}
{}\quad & \Tr_{\rm ad} F^2 = (N-2)\tr F^2\cr
        & \Tr_{\rm ad} F^4 = 3(\tr F^2)^2+(N-8)\tr F^4\cr
\noalign{\vskip.2cm}
{}\quad & \Tr_{\rm spinor} F^2 = {{\rm d}\over 8}\tr F^2\cr
        & \Tr_{\rm spinor} F^4 = {{\rm d}\over 4}
                    \Bigl[{3\over16}(\tr F^2)^2-{1\over4}\tr F^4\Bigr]}
}
where ${\rm d}=2^{{N\over2}-1}$ ($N$ even) and
${\rm d}=2^{{N-1\over2}}$ ($N$ odd) is the dimension of the spinor
representation.

$\underline{E_6}$
\eqn\esix{
\eqalign{
\Tr_{\ul{27}}F^2&\equiv 3\tr F^2\cr
\Tr_{\ul{27}}F^4&={3\over4}(\tr F^2)^2\cr
\noalign{\vskip.2cm}
\Tr_{\ul{78}}F^2&=12\tr F^2\cr
\Tr_{\ul{78}}F^4&={9\over2} (\tr F^2)^2}
}

$\underline{E_7}$
\eqn\eseven{
\eqalign{
\Tr_{\ul{56}}F^2&\equiv 6\tr F^2\cr
\Tr_{\ul{56}}F^4&={3\over2}(\tr F^2)^2\cr
\noalign{\vskip.2cm}
\Tr_{\ul{133}}F^2&=18 \tr F^2\cr
\Tr_{\ul{133}}F^4&=6 (\tr F^2)^2}
}

$\underline{E_8}$
\eqn\eeight{
\eqalign{
\Tr_{\rm ad}F^2&=30\tr F^2\cr
\Tr_{\rm ad}F^4&=9(\tr F^2)^2}
}

\listrefs
\end